\documentclass{emulateapj}
\accepted{June 29, 2016}

\shortauthors{Kesseli et al.}
\shorttitle{Models of Photometric Variability in Young Stellar Objects}

\begin{document}

\title{A Model for (Quasi-)Periodic Multi-wavelength Photometric Variability in Young Stellar Objects}

\author{Aurora Y. Kesseli\altaffilmark{1}, Maya A. Petkova\altaffilmark{2}, Kenneth Wood\altaffilmark{2}, Barbara A. Whitney\altaffilmark{3,4}, L. A. Hillenbrand\altaffilmark{5}, Scott G. Gregory\altaffilmark{2}, J.R. Stauffer\altaffilmark{6}, 
M. Morales-Calderon\altaffilmark{6}, L. Rebull\altaffilmark{6}, S. H. P. Alencar\altaffilmark{7}}

\altaffiltext{1}{Boston University, 725 Commonwealth Ave, Boston, MA 02215}

\altaffiltext{2}{SUPA, School of Physics \& Astronomy, University of St Andrews, 
North Haugh, St Andrews, Fife, KY16 9AD}

\altaffiltext{3}{Department of Astronomy, University of Wisconsin-Madison, 475 N. Charter St, 
Madison, WI 53706}

\altaffiltext{4}{Space Science Institute,  4750 Walnut St. Suite205, Boulder, CO 80301}

\altaffiltext{5}{Astronomy Department, California Institute of Technology, Pasadena, CA 91125, USA}

\altaffiltext{6}{Spitzer Science Center, California Institute of Technology, CA 91125}

\altaffiltext{7}{Departamento de F\'{\i}sica -- ICEx -- UFMG, Av. Ant\^onio Carlos, 6627, 30270-901, Belo Horizonte, MG, Brazil}

\email{aurorak@bu.com}

\begin{abstract}

We present radiation transfer models of rotating young stellar objects (YSOs) with hotspots
in their atmospheres, inner disk warps and other 3-D effects in the nearby circumstellar
environment.  Our models are based on the geometry expected from
the magneto-accretion theory, where material moving inward in the disk flows along magnetic field lines
to the star and creates stellar hotspots upon impact. Due to rotation of the star and magnetosphere, 
the disk is variably illuminated.  We compare our model light curves to data from the Spitzer
YSOVAR project \citep[e.g.][]{morales11, cody14} to determine if these processes can explain 
the variability observed at optical and mid-infrared wavelengths in young stars.
We focus on those variables exhibiting ``dipper" behavior that may be periodic,
quasi-periodic, or aperiodic.
We find that the stellar hotspot size and temperature affects the optical and near-
infrared light curves, while the shape and vertical extent of the inner disk warp 
affects the mid-IR light curve variations.
Clumpy disk distributions with non-uniform fractal density structure produce more stochastic light curves.
We conclude that the magneto-accretion theory is consistent with certain aspects of the 
multi-wavelength photometric variability exhibited by low-mass YSOs.
More detailed modeling of individual sources can be used to better determine the stellar hotspot and
inner disk geometries of particular sources.

\end{abstract}


\maketitle

\section{Introduction}
\begin{figure}[h,t!]
\begin{center}
\includegraphics[scale=0.45]{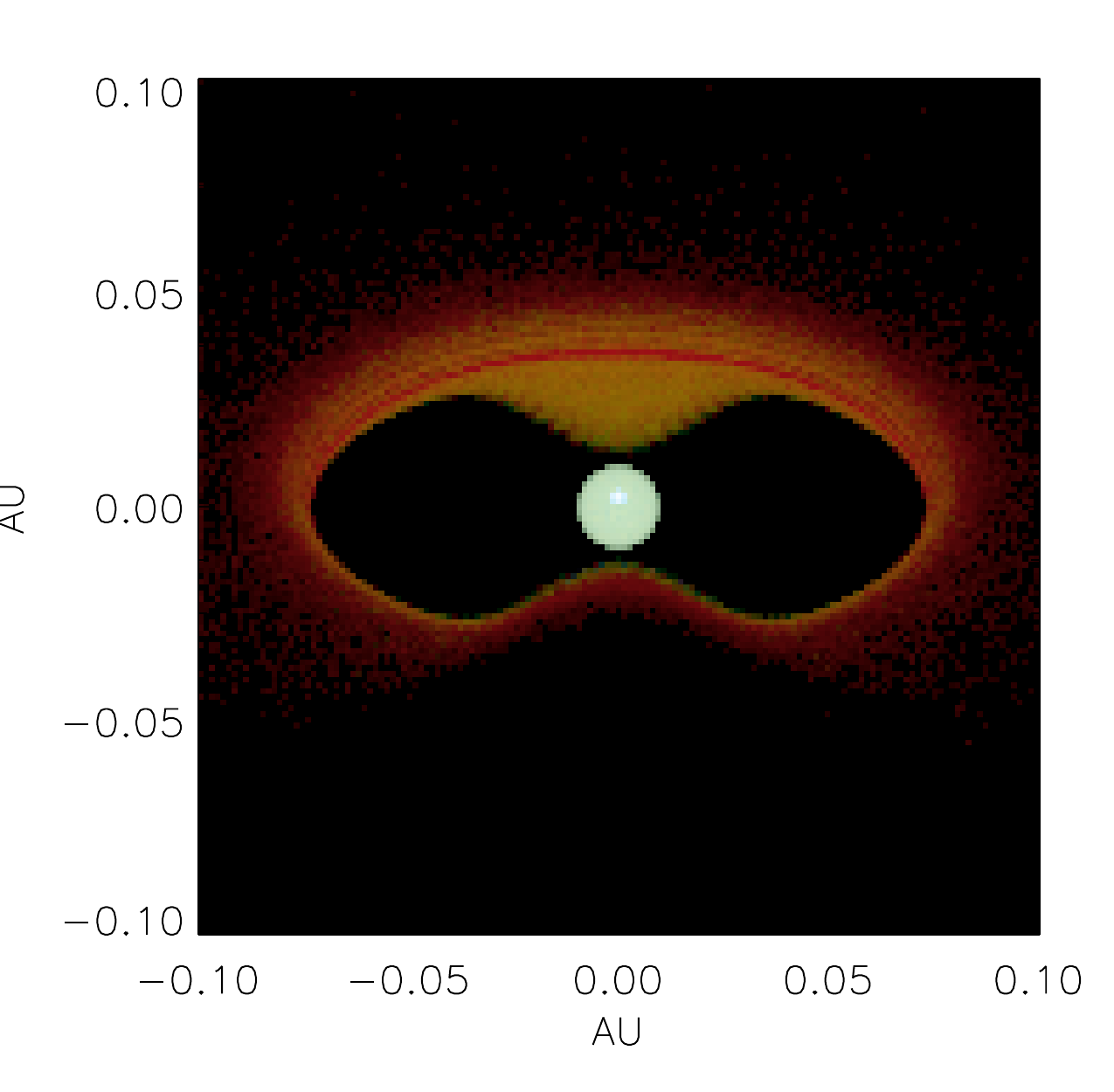}
\includegraphics[scale=0.45]{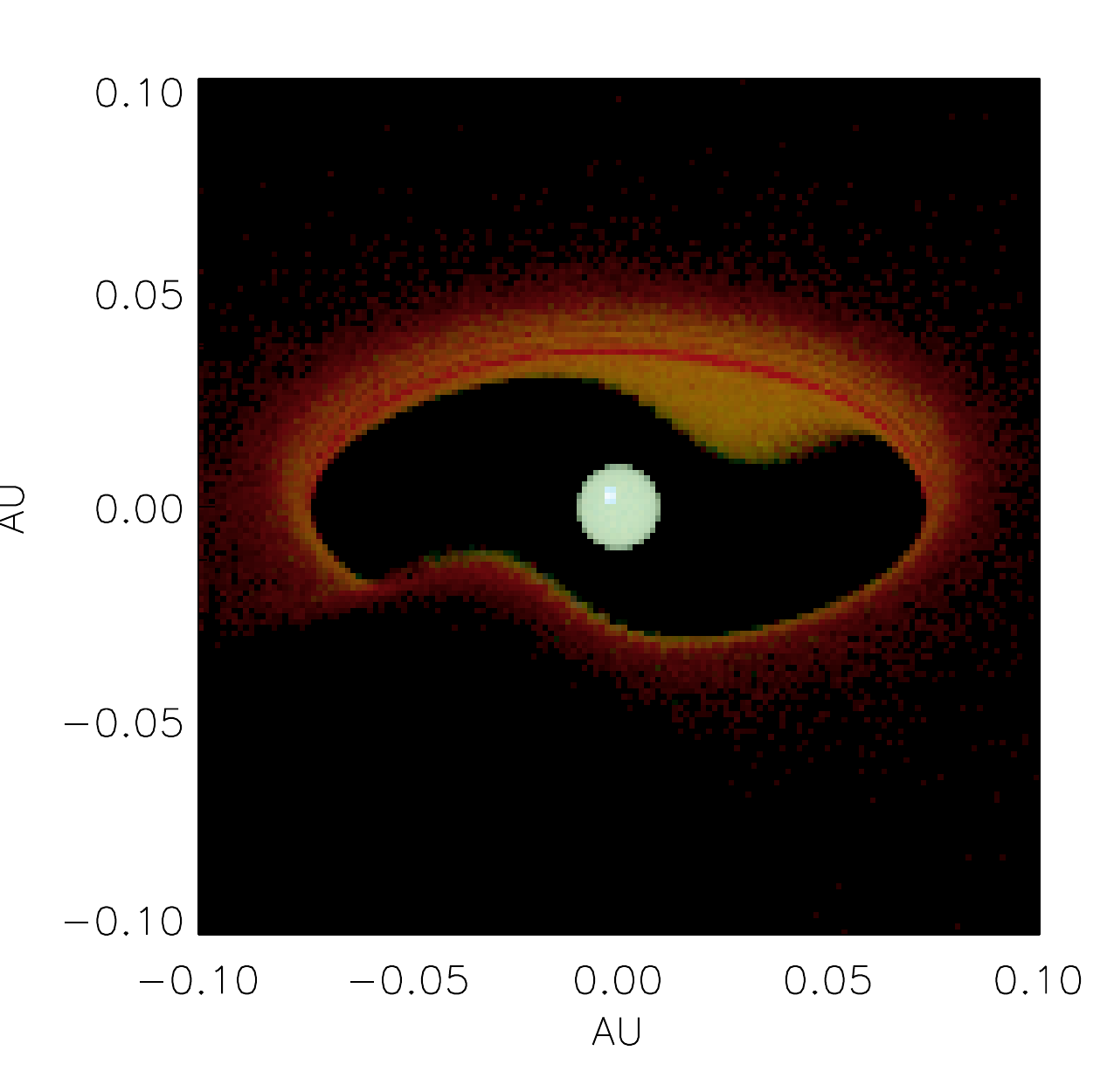}
\caption{\small 
3-color plots of the star and inner disk region represented by Model 1 
having two hotspots due to accretion that illuminate a truncated disk with an
inner warp (see Table 1 for details).  The disk is inclined 
at a viewing angle $i=60\arcdeg$, and is shown at azimuthal angles 
$\phi$= 0\arcdeg, and 20\arcdeg.
The color scale places V band (0.55 $\mu$m) as blue, J band (1.2 $\mu$m) as green and IRAC 4.5 $\mu$m as red.}
\label{f:image}
\end{center}
\end{figure}

Multi-wavelength studies of the variability of young stellar objects (YSOs) probe the combined stellar 
and circumstellar properties of newly forming stars along with angular momentum driven phenomena such as 
stellar rotation and binary orbital motion.
Optical and near-infrared data are sensitive to the stellar photosphere (hot and cool spots), and 
other energetically ``hot" regions (accretion
columns, chromospheres), as well as scattering from the circumstellar material.
Observations at mid-IR and longer wavelengths offer a
new perspective as they are sensitive to variability associated with
``warm" or ``cool" regions --- the disks and envelopes of YSOs.   
Figure  \ref{f:image} illustrates that different wavelengths dominate different regions by showing a 3-color plot of 
one of our models of a spotted star surrounded by a warped accretion disk.

The optical variability of accreting YSOs has been successfully interpreted in the context of 
the magnetospheric accretion model.
In this model, the inner disk is truncated, and material flows from the disk to the star along
stellar magnetic field lines \citep{ghosh78,konigl91}.
As the free-falling material reaches the star, the kinetic energy is dissipated in shocks 
at the stellar surface \citep{konigl91}.
The stellar magnetic field is often inferred to not be aligned with the rotation axis based on line emission modeling \citep{donati11}, resulting in photometric
modulation as the shock columns move in and out of view \citep{mahdavi98, gregory11}.
Strong H$\alpha$ (and other) line emission and blue excesses are produced by the inflowing gas and shock columns 
\citep{hartmann94,gullbring98,muzerolle01}.
The lightcurves of accreting YSOs show variations on a variety of timescales
and with a variety of color-magnitude effects \citep{Herbst94}. 
Timescales on the order of a few hours
track material in free-fall from the inner disk to the stellar surface.  The time for an inner disk asymmetry
to transit the stellar surface is $\sim 0.3$ days on average. 
The stellar rotational modulation is typically $\sim1-8$ days \citep{rebull04}.
Disk accretion rates and magnetospheric structure changes occur on timescales of days to weeks to years.
The color variability ranges from essentially colorless amplitude variability, indicating achromatic 
or ``black" processes, to large color variability, indicating substantial changes in accretion or extinction.

\citet{morales11} and \citet{cody14} presented results of multi-wavelength photometric 
monitoring of the Orion Nebula Cluster (ONC) and NGC 2264, as part of the young stellar object variability (YSOVAR) 
project that also includes many smaller clusters, as summarized in \citet{rebull14}.
Among thousands of YSOs, 70\% of those with mid-IR excess are variable at levels typically 0.1 to 0.2 magnitudes but some have amplitudes as high as 0.5 mag. 
The YSOs observed exhibit many different behaviors, but can be grouped into a few main categories based on light curve morphology: 
periodic/quasi-periodic, dippers (both periodic and irregular), bursters/accretors
(almost always irregular), stochastic variables, and stars showing either brightening or fading trends covering the full duration of the time series.  

The periodic light curves can have relatively symmetric and regular flux variations, 
but there is also a sub-class of the periodic sources with asymmetric lightcurves that show pronounced ``dips."
Other light curves exhibit quasi-periodic behavior, with additional upward or downward 
trends in brightness that render them not detected as significantly periodic under Fourier analysis
though with semi-ordered and repeated variations diagnosed using the ``Q" statistic of \citet{cody14}.   
Like the periodic sources, the quasi-periodic objects  may be roughly symmetric in their brightness variations, 
or with pronounced ``dips."  Such ``dipper'' sources may be periodic with regular dips in brightness, 
quasi-periodic as described above, or irregular with dips occurring 
much more stochastically relative to a defined stable flux level.  
An obvious physical interpretation for this category is variable extinction, but we also propose an alternate model
based on variable illumination.

Another YSOVAR category is the inverse of the dippers, the ``bursters", that are characterized by flux bursts and excess brightness peaks 
on various time scales, with mostly constant flux otherwise, and irregular repetition. 
A popular interpretation for this category is variable accretion.
Some light curves are neither periodic nor quasi-periodic 
but exhibit large and/or small, stochastic, brightness variations 
over a few days, possibly due to a combination of extinction and accretion events \citep{cody14,stauffer15}. 
The ``trender" category is likely dominated by processes occurring outside the magnetospheric region,
where the dynamical time scales are longer than the few days to week long variations that typify the other categories.

In this paper we present models intended to apply only to the various forms of periodic and quasi-periodic
lightcurves, especially those of the ``dipper" variety.
Periodicity naturally arises from the rotation of the star and Keplerian rotation within the disk.
We illustrate how variations in accretion properties and inner disk geometry affect the brightness, 
including wavelength-dependent effects, which can be used to infer the physical processes 
responsible for the observed variations due to stochastic accretion. 
In section 2 we describe the star-hotspot-accretion disk geometry we adopt for our radiation transfer models. Section 3 presents 
the photometric and polarimetric variability from our models. In Section 4 we compare our models to observations and we summarize our 
findings in Section 5.

\section{Accretion Disk Models}

We use a Monte Carlo radiation transfer code \citep{whitney03, whitney04, whitney13} to create models of young stellar objects. The code utilizes a purely geometric model of dust radiation transfer and does not include any magneto-hydrodynamics. Our code computes the 
emergent spectral energy distribution and multi-wavelength images (including polarization arising from scattering off dust grains) for a dusty disk plus envelope heated by starlight and accretion luminosity. 
We have modified the code of \citet{whitney03} to include stellar hotspots, warped inner disks, fractal clumping, spiral arms, and other 2-D and 3-D features.  
The equations describing the accretion model, hotspot, and disk geometries, are described in detail in \citet[Section 3.8]{whitney13}. The model does not account for possible magnetic field grain alignment effects, as discussed in e.g. \citet{cho07}.
We do not include the emission from gas inside the dust destruction radius, only the star and dust emission. 
In what follows we present models for a range of accretion rates, stellar hotspot parameters, and the
shape and location of the inner edge of the dust disk.

We consider a typical low mass classical T~Tauri star (CTTS) having $M_\star = 0.5M_\odot$, $T_\star = 4000$~K, and $R_\star = 2R_\odot$ that is surrounded by an accretion disk of mass $M_{\rm disk} = 0.05M_\odot$ and outer radius 100~AU. We assume that the star and inner disk are locked with the same rotation period due to the angular momentum lost from outflows, for example, accretion-powered winds \citep{mattpudritz2005}, extended disk winds \citep{ferreira2000}, X-winds \citep{mohantyshu2008} launched from the star-disk interaction region, or magnetospheric ejections \citep{aarnio2012,zanni2013}. The inner disk radius is generally set to be the dust sublimation radius for this typical cTTS using the formula from \citet{whitney04}, $R_{\rm sub} = (T_{\rm sub} / T_\star) ^{-2.085} R_\star$, with $T_{\rm sub} = 1600$~K, but we also explored some models where the inner disk radius was set to three and five times this value (Table \ref{t_models}). While such models are unlikely to be disk-locked, we include them as an initial exploration of parameter space. 
The disk is slightly flared with a scale height that depends on radius as $h(r) = h_0 (r/R_\star)^\beta$. We adopt $h_0 = 0.01\, R_\star$, $\beta = 1.25$ giving a scale height $h(100\, {\rm AU}) \approx 10$~AU. 

In the magnetospheric accretion model, material from the accretion 
disk flows onto the surface of the star following magnetic field lines. In a stable model, a slightly tilted large-scale magnetosphere truncates the disk, and the in-falling material creates two ordered hotspots separated by $180^\circ$ in azimuth where the flow hits the surface of the star \citep{romanova08}. 
We set the size of the hotspot to be the median size estimated by \citet{gullbring98} of 0.7\% 
of the surface area of the star. For a spot temperature of $10^4$~K this gives an accretion rate of $9.21\times10^{-8} M_\odot\, {\rm yr}^{-1}$, where we use 
equations 4 to 7 from \citet{whitney13} relating spot size, temperature, and accretion rate. 
While this accretion rate of our initial model is higher than typical cTTS \citep{manara14, ingleby14, hh08}, and would require a dipole field larger than has currently been observed in cTTS, we find that models with lower accretion rates 
reduce the amplitude but do not alter the shape of the variability in our simulations (e.g., see Models 3 and 4 in Table 1 and Figure 2). 
The star-spot temperature contrast may be larger than in some of our models because accretion hotspots are often found within large cool spots \citep{donati07}.  We do not include cool spots in our models, however future model developments should explore their inclusion. \textbf{\citet{venuti15} estimate that $\lesssim 10\%$ of observed light curves in the actively accreting (defined by observed UV excesses) cTTS sample are dominated by cool spots. In systems with accretion, we expect the effect of cool spots to be less dominant since their amplitudes are only around 0.1 magnitude in the $r$ band \citep{cody14}.} The contrast in optical is much higher than in the infrared bands \citep{cody14}, so the variability pattern in the IRAC bands will be dominated by the circumstellar effects we have modeled.

The hotspot in the initial model is at 45\arcdeg\ latitude and emits 37\% of the total luminosity from the star.  The mid-latitude spot is motivated by modeling observations of line emission on T Tauri stars \citep[e.g.,][]{donati10}. \textbf{We also include some models with higher latitude hotspots, as an increasing number of T Tauri stars modeled using the line emission technique show this configuration \citep{donati10, donati12, donati13}. }At the inner edge of the disk the dust sublimates, 
so the material flowing
onto the hotspots is gaseous and assumed to be optically thin and so has no effect on the radiation transfer of stellar radiation. We modify our disk surface to include warps at the same longitude as the stellar hotspots, where the dust is uplifted with the gas as it flows towards the star. Our description of a disk ``warp" is 
an azimuthal variation of the disk scale height as shown in Figure 1 and described by equation 8 in \citet{whitney13}.
See \citet{romanova13} for dynamical models of warped disks. 
In order to simulate an unstable accretion disk where material penetrates the magnetosphere to reach the star at
lower latitudes than in the stable case, we used a fractal generating algorithm to create a clumpy inner 
disk, with the amount of clumped to smooth matter greater than 25\% \citep[see][Section 3.7]{whitney13}. Similar models are used to demonstrate that when unstable accretion occurs, gas flows onto the stellar surface at many locations, uplifting dust and creating a clumpy disk \citep[see][Figure 1]{romanova08}. 

Table \ref{t_models} shows the parameters for the different models presented in this paper. The fractional area of the hotspots is the percent of the total surface area of the star that the hotspot covers, 
and the temperature of the hotspot is dependent on this area so as to maintain a fixed accretion rate with a smaller hotspot area requiring a higher hotspot 
temperature. The accretion rate is the global accretion rate, which is calculated from the combination of accretion luminosity liberated in the disk and on the 
stellar surface due to material from the disk that is impacting the star at the hotspots \citep[see description of accretion luminosities in][equations 4 to 7]{whitney13}. 
The exponent of the disk warp is the parameter $w$ in equation 8 of \citet{whitney13}. This parameter 
affects the width of the disk warp (smaller exponent corresponding to larger warp) and hence the shape 
of the light curve. The fractal clumping ratio is the ratio of clumped to smooth material 
in the disk.  Most of our models do not utilize the fractal clumping parameter and for those that do, a 
clumpier disk produces a more stochastic light curve.  For the dust within the disk we follow our previous work and adopt two dust models, one has a size distribution representative of small grains in the 
interstellar medium \citep{kim94} while the second 
extends to larger grain sizes \citep[see][Table 1, Model 1]{wood02}. The large-grain model has a smaller scaleheight, thus approximating grain growth and settling within the disk \citep{dullemonddominik2004}. 
The latitude of the spots gives the angle measured from the edge-on viewing angle to the spots (if they are symmetrical). If there are multiple spots, or spot regions the latitudes that they cover are stated. Lastly, a short description is included for each model. The inclination angle parameter is also changed, however, we do not include this in the table because some of the models are shown at multiple inclinations. The inclination angles are instead stated in the heading of each sub-figure, where an inclination angle (i) of 90\arcdeg\ is edge-on and 0\arcdeg\ is face-on.

\begin{deluxetable*}{c c c c c c c c l}
\tablecolumns{9}
\centering
\tablewidth{0pt}
\tablecaption{Model Parameters} 

\tablehead{\colhead{ } & \colhead{Number} & \colhead{Fractional} & \colhead{Accretion} & \colhead{Exponent} & \colhead{Inner Disk} & \colhead{Fractal} & \colhead{Latitude} &\colhead{Model} \\
\colhead{ } & \colhead{of} & \colhead{Area of} & \colhead{Rate} & \colhead{of Disk} & \colhead{Radius} & \colhead{Clumping} & \colhead{of Spots} & \colhead{Description} \\
\colhead{ } & \colhead{Hotspots} & \colhead{Hotspots} & \colhead{(M$_{sun}$/yr)} & \colhead{Warp} & \colhead{(AU)} & \colhead{Ratio} & \colhead{(degrees)} & \colhead{}}

\startdata
model 1 &2& 0.7\% & $9\times 10^{-8}$ & 81 & 0.06 &-&45&initial model\\ 
model 2 &2& 0.7\% & $3\times 10^{-8}$ & 81 & 0.06 &-&45&reduced accretion\\ 
model 3 &2& 30\% & $9\times 10^{-10}$ & 81 & 0.06 &-&45&large hotspots\\ 
model 4 &2& 30\% & $9\times 10^{-10}$ & 5 & 0.06 &-&45&large hotspots; large warp\\ 
model 5 &2& 0.7\% & $9\times 10^{-8}$ & 5 & 0.06 &-&45&large warp\\ 
model 6 &2& 0.7\% & $9\times 10^{-8}$ & - & 0.06 &-&45&no warp in inner disk\\ 
model 7 &1& 0.7\% & $9\times 10^{-8}$ & 81 & 0.06 &-&45&single hotspot\\ 
model 8 &2& 0.7\% & $9\times 10^{-8}$ & 81 & 0.31 &-&45&larger inner disk radius\\ 
model 9 &2& 0.7\% & $4\times 10^{-9}$ -- $6\times 10^{-8}$& 81 & 0.19 &-&45&variable accretion\\ 
model 10 &2& 2\% & $9\times 10^{-8}$ & 81 & 0.06 & 0.25 &45&low clumpiness disk\\ 
model 11 &2& 2\% & $9\times 10^{-8}$ & 81 & 0.06 & 0.5 &45&moderate clumpiness\\ 
model 12 &2& 0.7\% & $9\times 10^{-8}$ & 81 & 0.06 &-&60&high spot latitude\\ 
model 13 &2& 0.7\% & $9\times 10^{-8}$ & 81 & 0.06 &-&80&higher spot latitude\\ 
model 14 &2& 0.7\% & $9\times 10^{-8}$ & - & 0.06&-&60&no warp, high spot latitude\\ 
model 15 &2 regions\tablenotemark{1} & 2.7\% &$10^{-8}$ &41&0.05&-&37-60&complex dipole hotspots\\ 
model 16 &4 regions\tablenotemark{1} & 3.0\% &$10^{-8}$ & 41 & 0.08 &-&0-10, 51-79&octupole hotspots\\ 
model 17 &many regions\tablenotemark{1}&4.7\%&$10^{-8}$&41&0.10&-&0-51&many hotspots\\ 
\tablenotetext{1}{See section 3.4 for details on the hotspot geometry of these models}

\enddata

\label{t_models}
\end{deluxetable*}

\section{Results}

To compare with observations, we construct light curves from our radiation transfer models at optical ($V$ and $I$ bands),
near-infrared ($J$ and $K$ bands) and {\it Spitzer Space Telescope's} Infrared Array Camera (IRAC) mid-infrared (3.6$\mu$m and 4.5$\mu$m) wavelengths. 
Figure \ref{f:image} shows multiwavelength images from one of our simulations that illustrates the geometry 
of the model. Only the star and the disk dust (no gas) appear in the image, as dust is assumed to be the dominant contributor to the 
opacity and the emissivity 
and hence to the continuum fluxes observed at the YSOVAR wavelengths.  

In our models, the emission within the IRAC bands arises from the heating of the inner disk wall, and is brightest when the projected area of the disk wall is largest.  This occurs when the photospheric hot spot is on the far side of
the star (not visible), heating up the back wall of the uplifted disk. 
The near side of the inner disk wall is mostly in shadow of the outer disk and is not seen when
illuminated by the hotspot.  At the shortest wavelengths,
the $V$, $I$, and $J$ light curves are dominated by the stellar hotspots as they pass in and out of view.  
The scattering contribution from the inner disk is small in comparison to direct light from the hotspot  (see Figure \ref{f:image}). The star is therefore brightest when the hot spot is facing the observer.
The near-infrared variations have behavior that is intermediate between the optical and IRAC mid-infrared, 
sometimes showing in-phase variation, little or no variation, and sometimes out-of-phase variation with IRAC, 
depending on the system inclination, inner disk radius, and stellar-to-hotspot temperature contrast.

In the following sub sections we present synthetic photometric and polarimetric lightcurves 
for a subset of our models that reproduce several broad categories of observed variability 
identified within YSOVAR.  We discuss periodic and quasi-periodic (sinusoidally varying) sources, then so-called dipper sources, 
and finally some aspects of the stochastic lightcurves that may be explained by variable and/or unstable accretion 
as in the models of \citet{romanova08}.

\subsection{Periodic and Quasi-Periodic Variations} 
Stars that display sinusoidal light curves are considered members of the periodic group. 
This group also includes stars with light curves that are not strictly periodic or 
do not have the same periodic nature in all bands, but do have an overall periodic trend \citep{cody14}. 
Sinusoidal variations at optical wavelengths can be explained with hotspots
on the star that rotate in and out of the field of view, while most of the variation in the mid-infrared IRAC data 
is due to the geometry of the warped disk and the variable heating and thermal radiation of the rotating inner disk wall.  

\citet{morales11} show many different examples of light curves in the ONC that they categorized as periodic.
In some cases there is a steady flux in the IRAC bands accompanying the periodic variations in the optical, 
while other cases 
show the opposite, with little variation out to the $J$ band and more variability at longer wavelength IRAC bands. 
There are also sources exhibiting variability at all wavelengths. 
 
\begin{figure*}[h,t!]
\begin{center}
\includegraphics[scale=0.6]{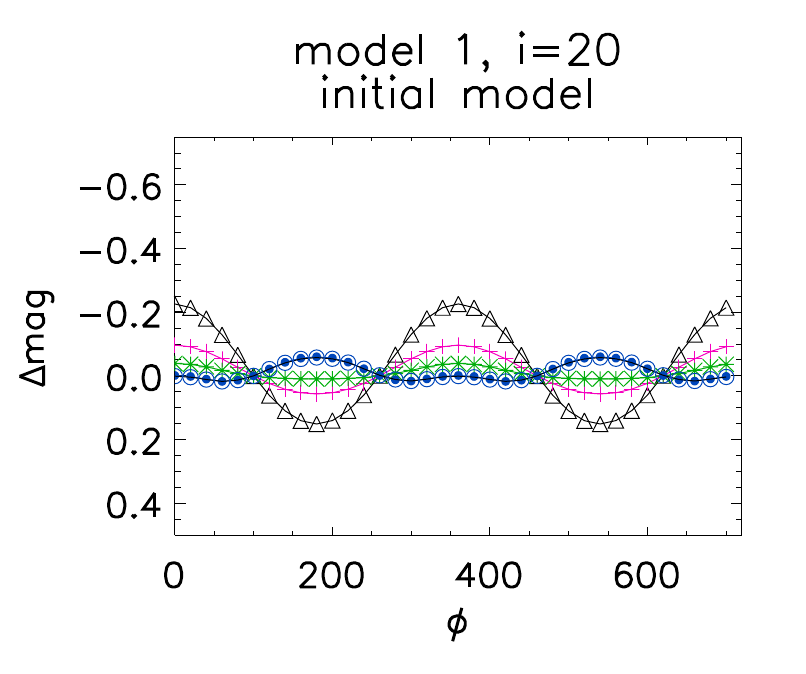}
\includegraphics[scale=0.6]{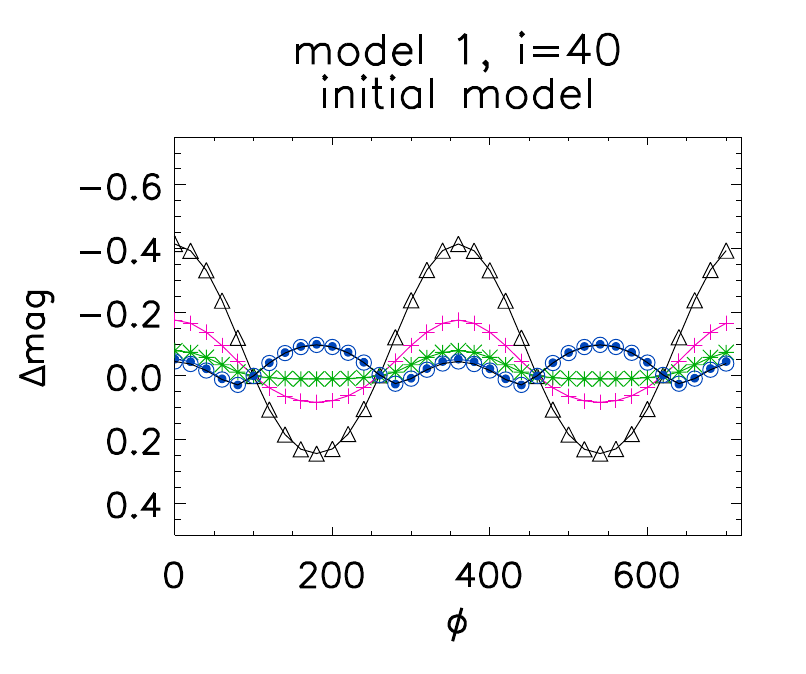}
\includegraphics[scale=0.6]{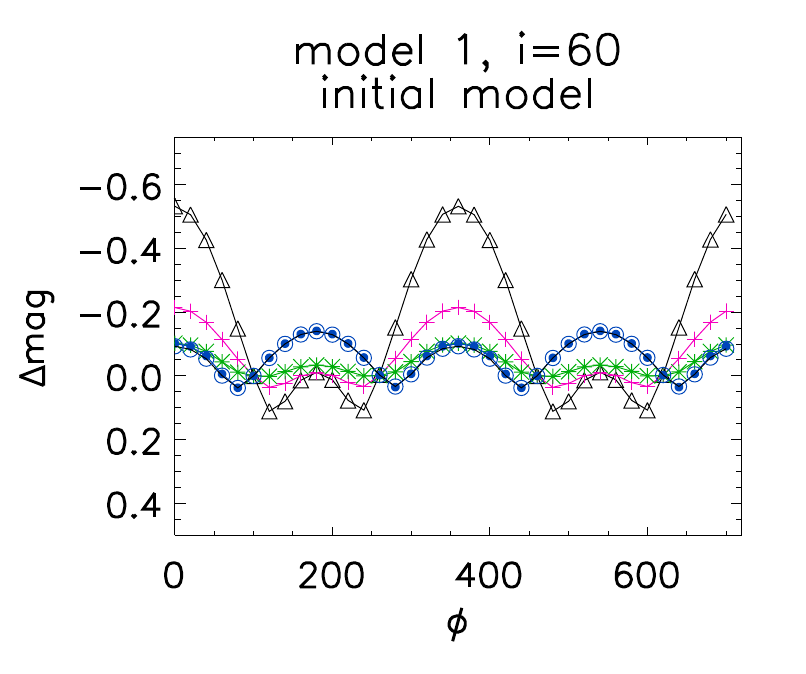} 
\includegraphics[scale=0.6]{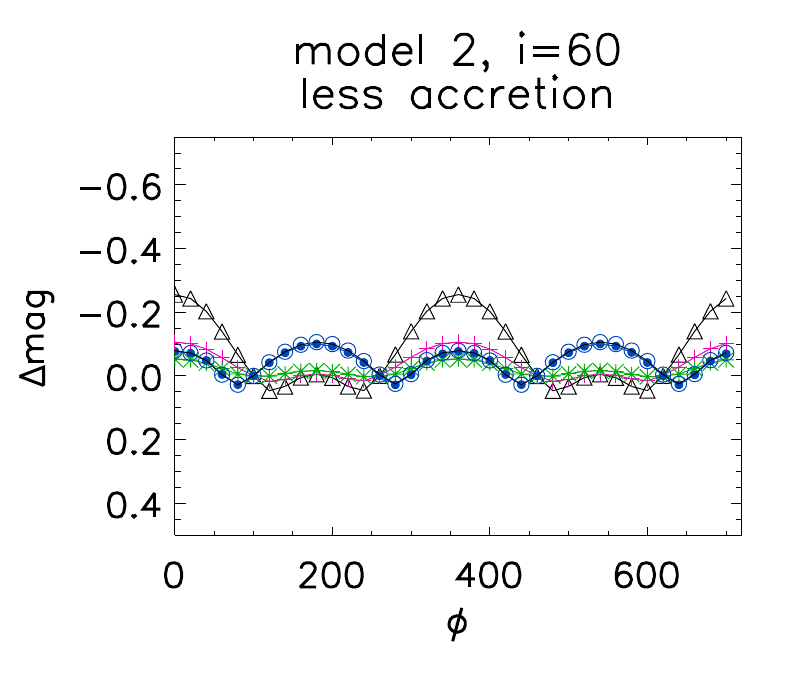}
\includegraphics[scale=0.6]{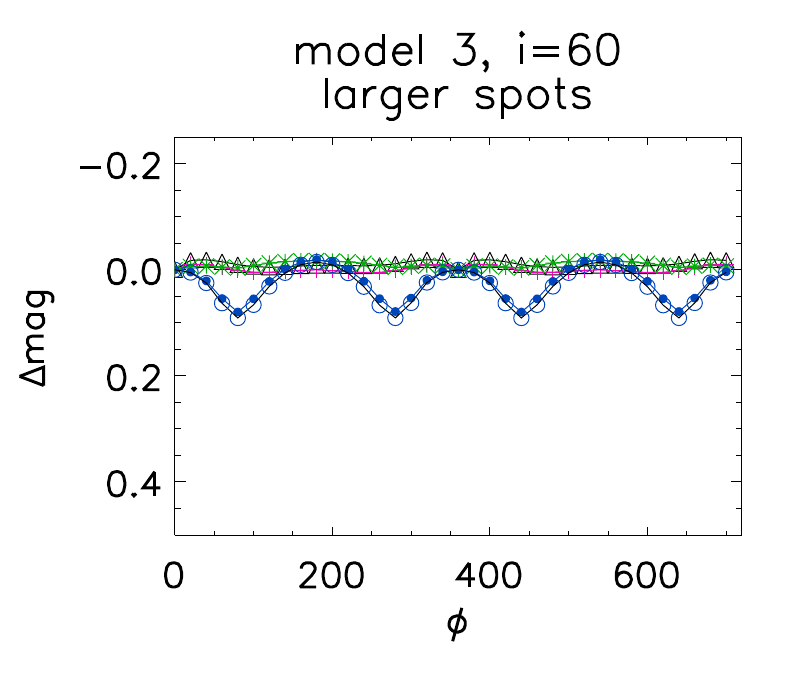}
\includegraphics[scale=0.6]{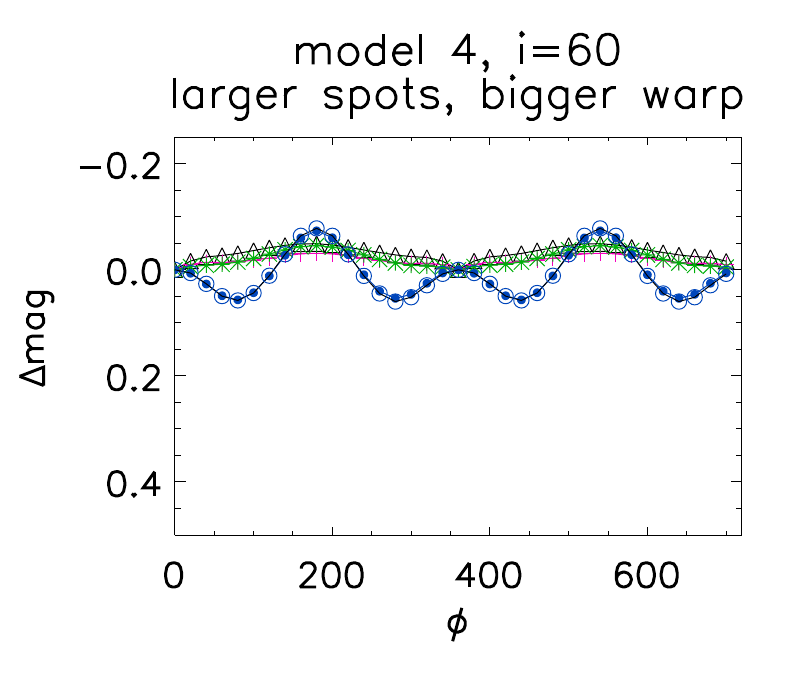}
\includegraphics[scale=0.6]{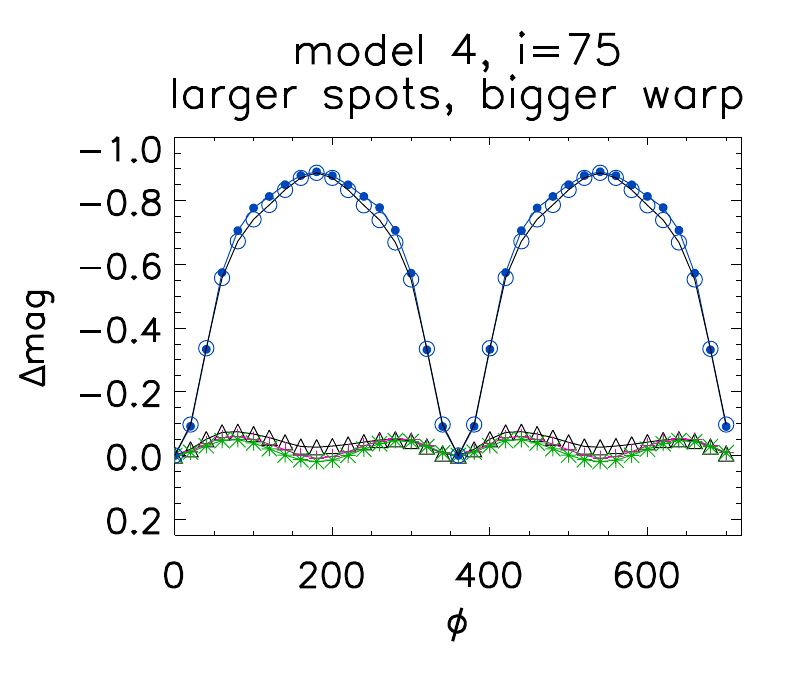}
\includegraphics[scale=0.6]{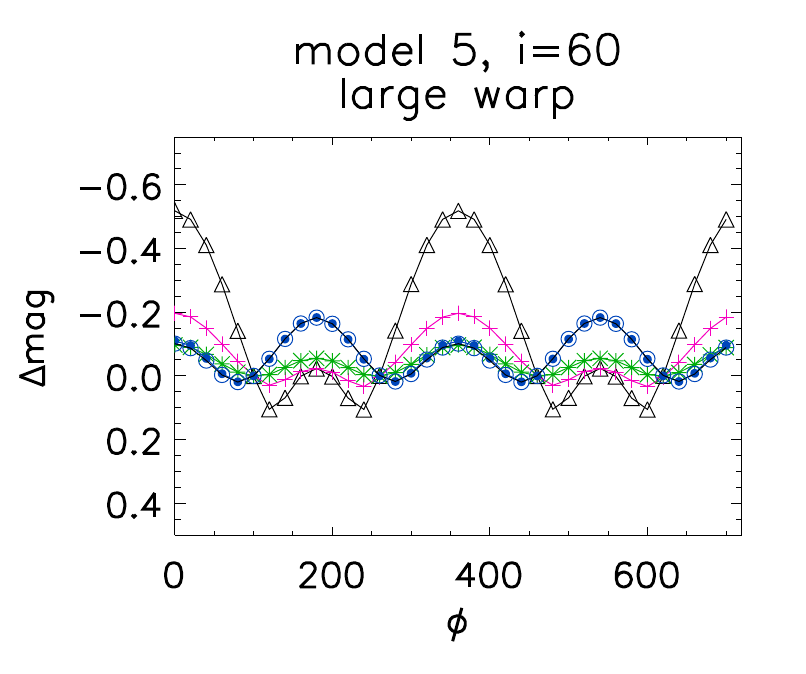}
\includegraphics[scale=0.6]{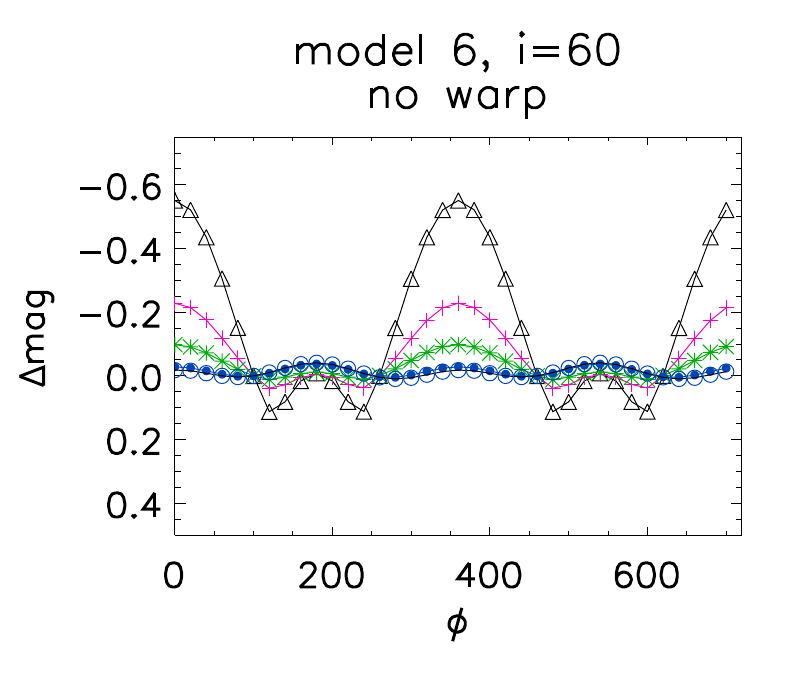}
\includegraphics[scale=0.6]{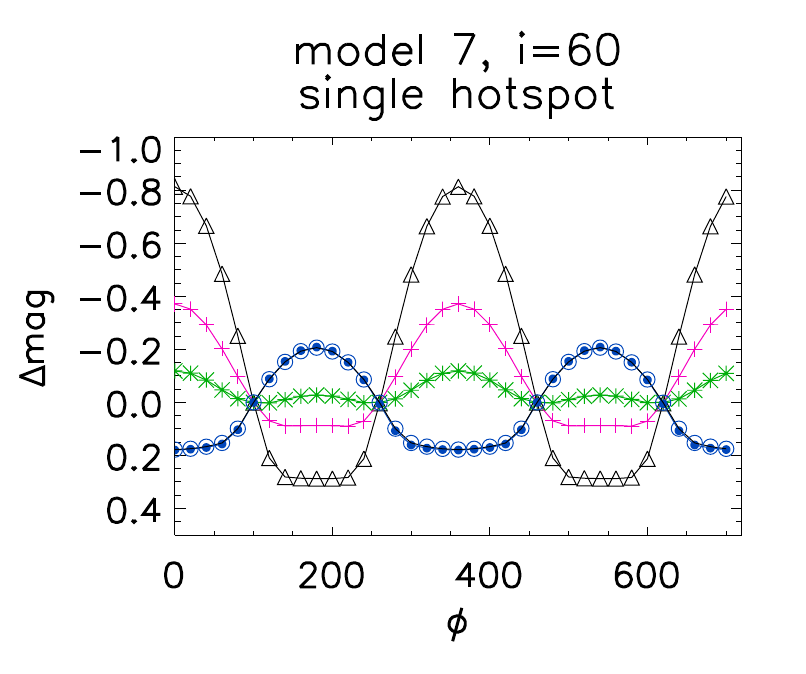}
\includegraphics[scale=0.6]{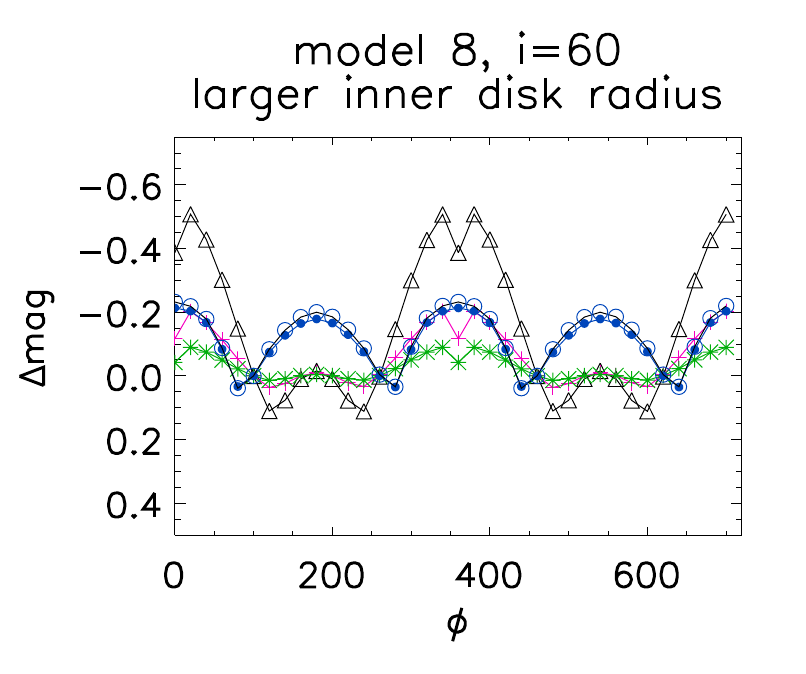}
\includegraphics[scale=0.6]{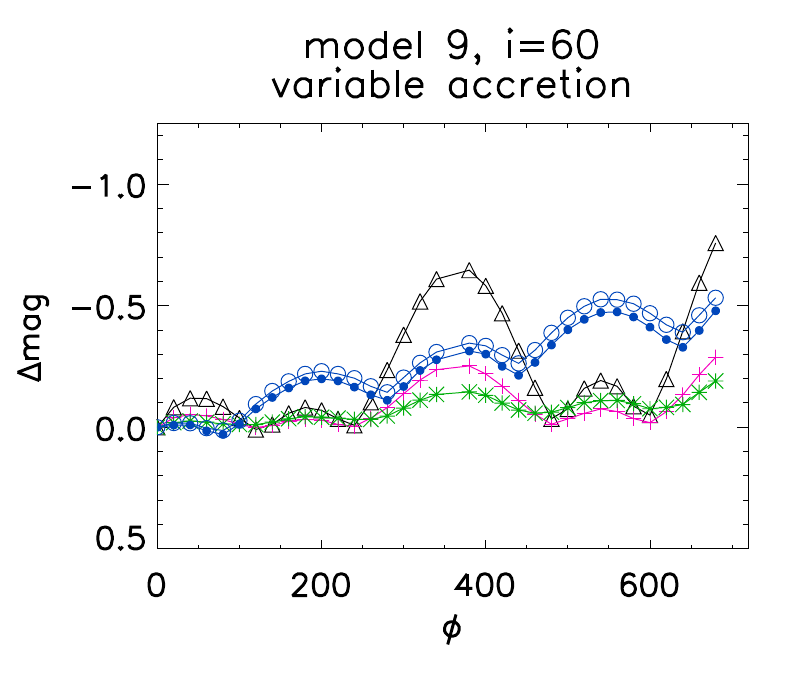}
\caption{\small 
Model light curves for the periodic category over two rotations; see Table 1 for detailed description.  
Some models are shown at multiple inclination values.
The symbols are chosen to correspond to the lightcurves illustrated 
in \citet{morales11} with the addition of black/triangles corresponding
to V-band, pink/plus: I-band, green/asterisk: J-band, 
blue/dot: IRAC [3.6], blue/circle:  IRAC [4.5].  
The magnitudes have been normalized separately in each band at either 0$^\circ$ or 100$^\circ$ for best readability.
}
\label{f:lc_period}
\end{center}
\end{figure*}

\begin{figure*}[h,t!]
\begin{center}
\includegraphics[scale=0.6]{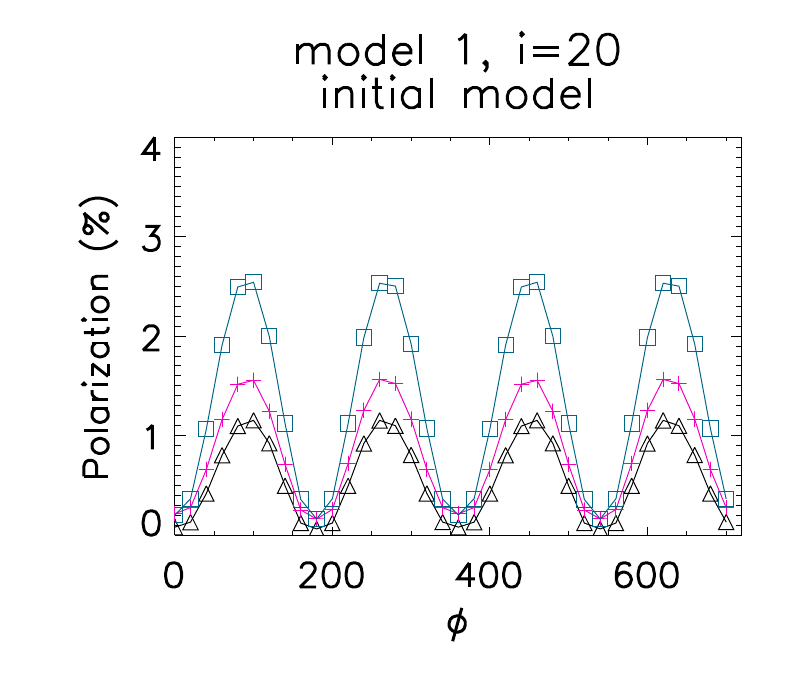}
\includegraphics[scale=0.6]{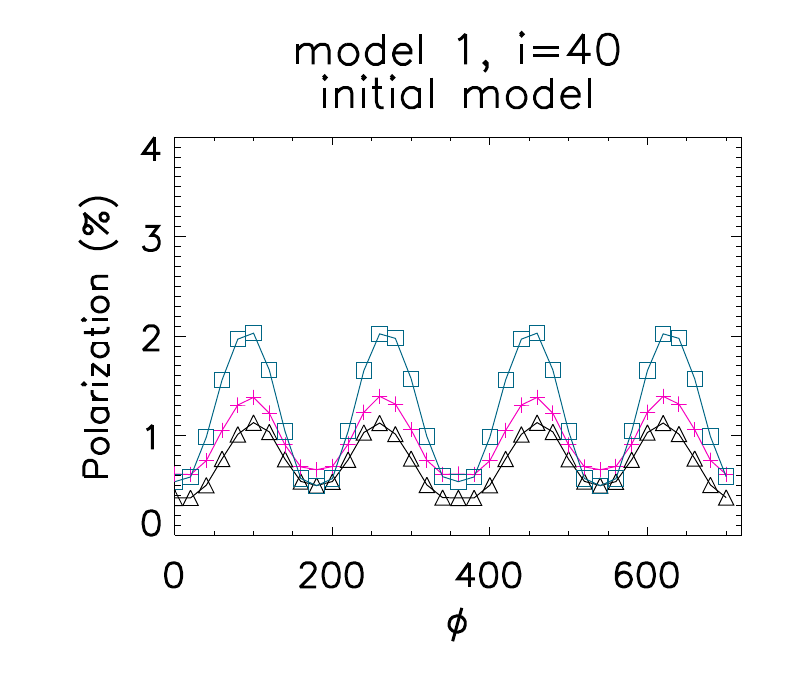}
\includegraphics[scale=0.6]{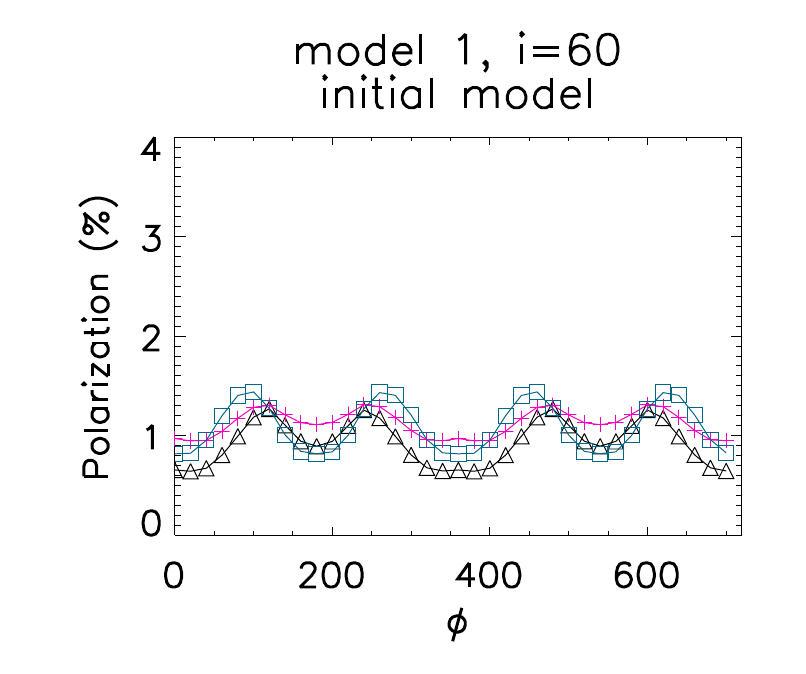}
\includegraphics[scale=0.6]{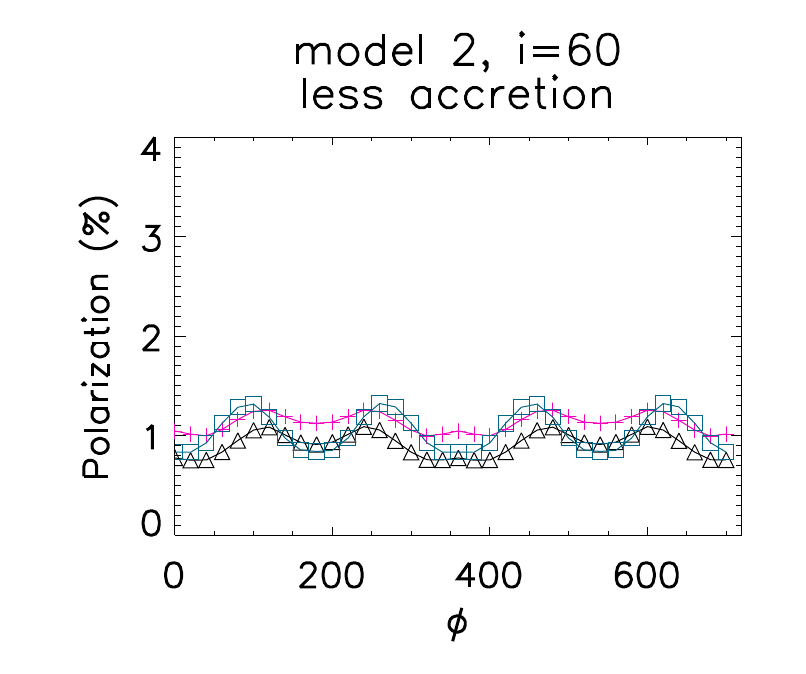}
\includegraphics[scale=0.6]{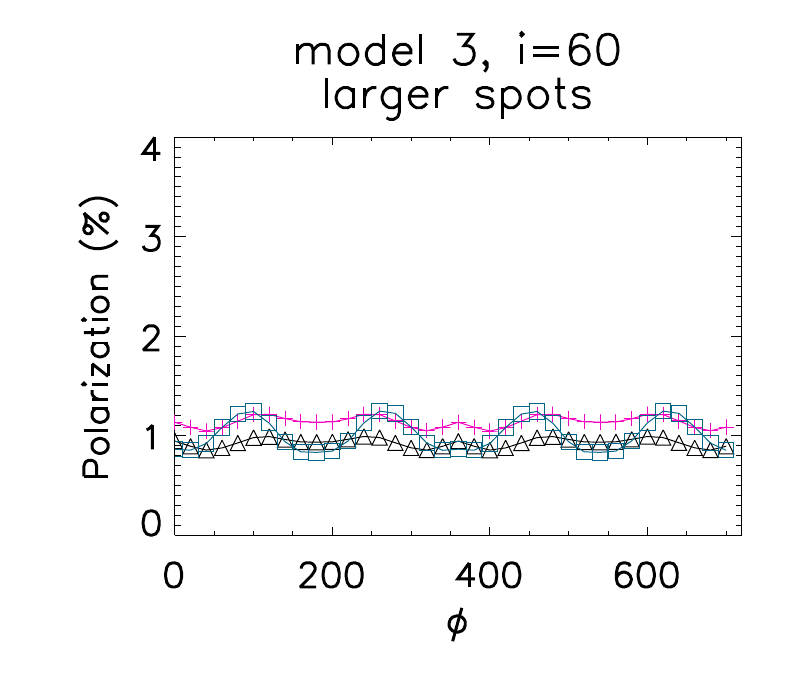}
\includegraphics[scale=0.6]{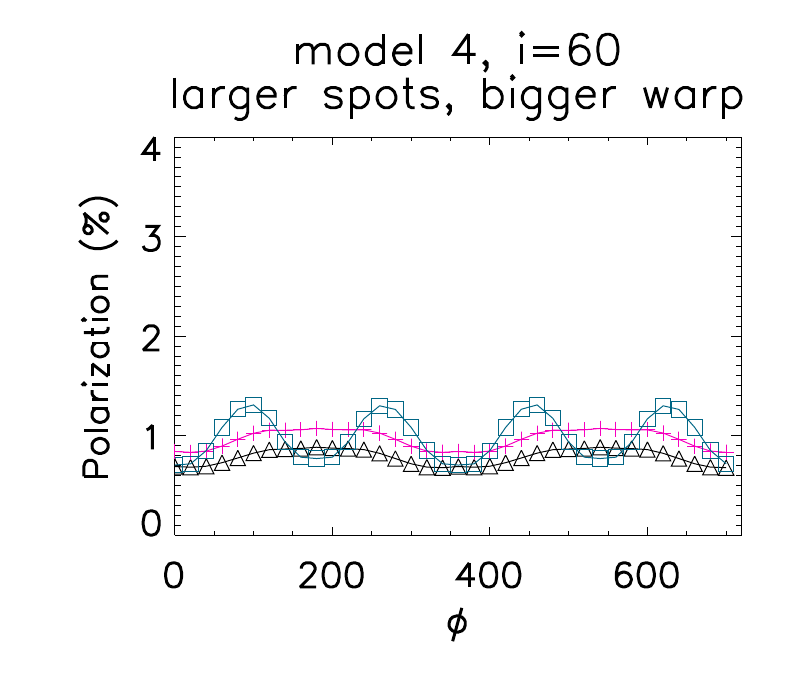}
\includegraphics[scale=0.6]{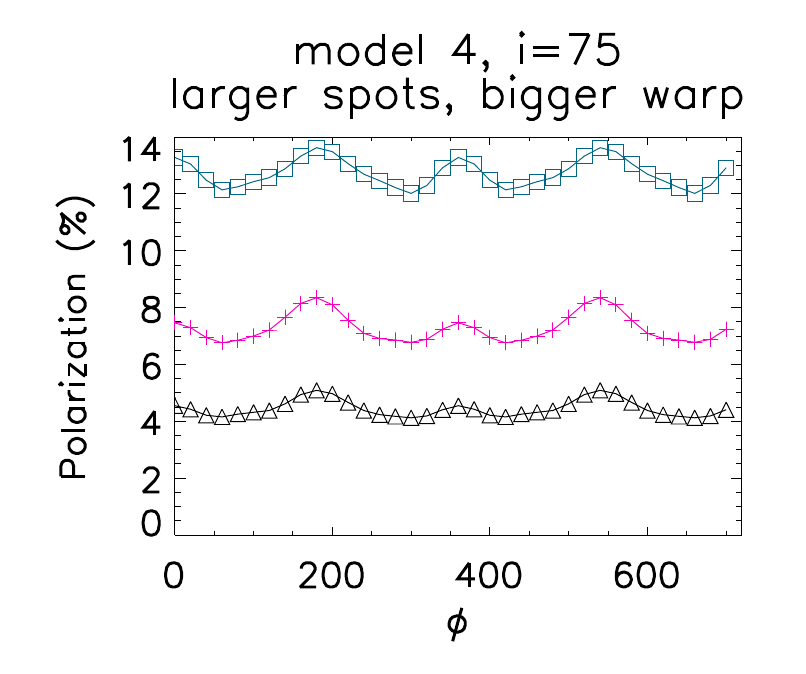}
\includegraphics[scale=0.6]{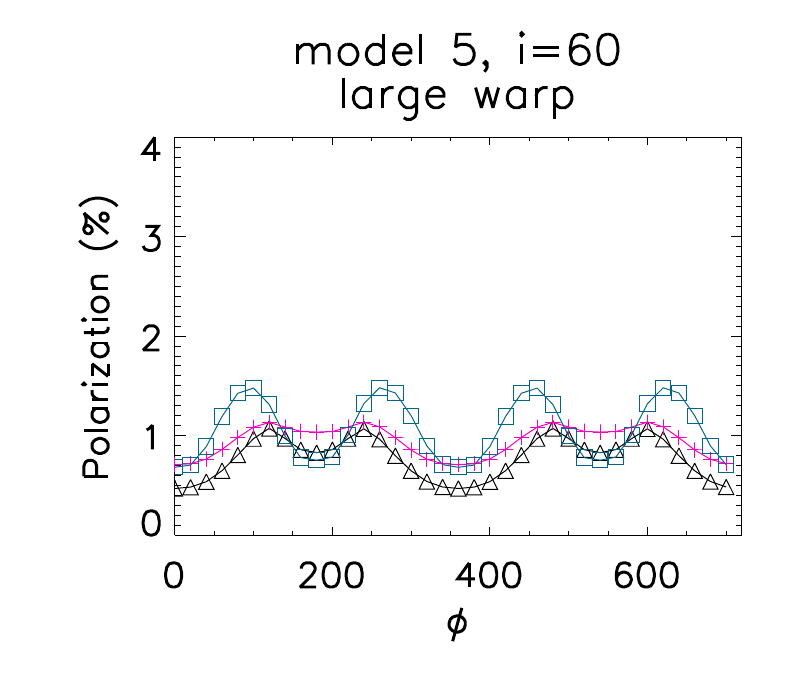}
\includegraphics[scale=0.6]{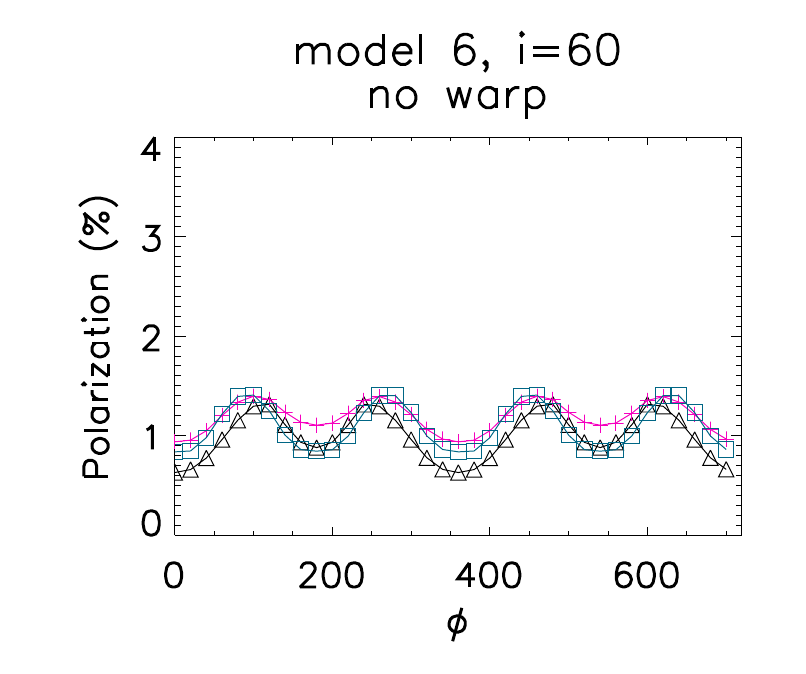}
\includegraphics[scale=0.6]{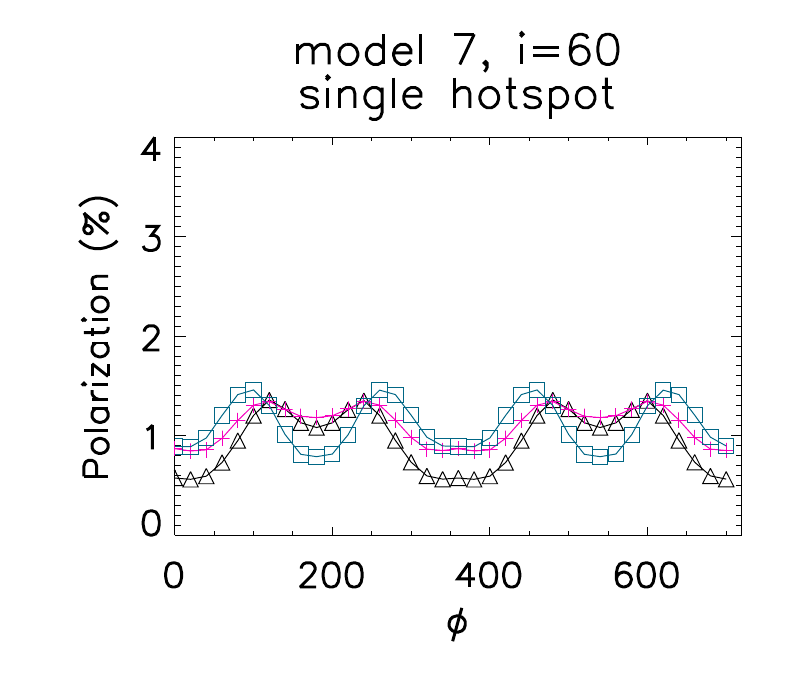}
\includegraphics[scale=0.6]{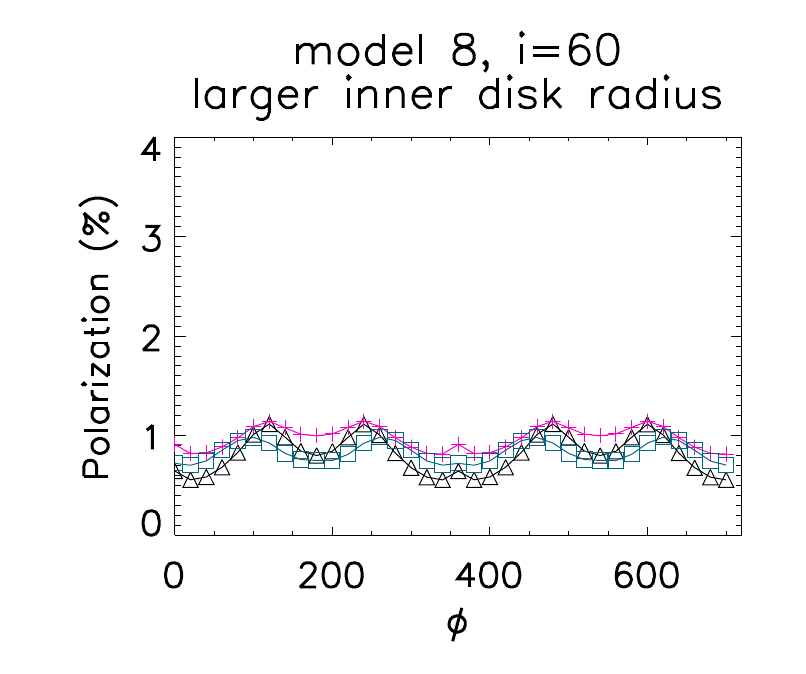}
\includegraphics[scale=0.6]{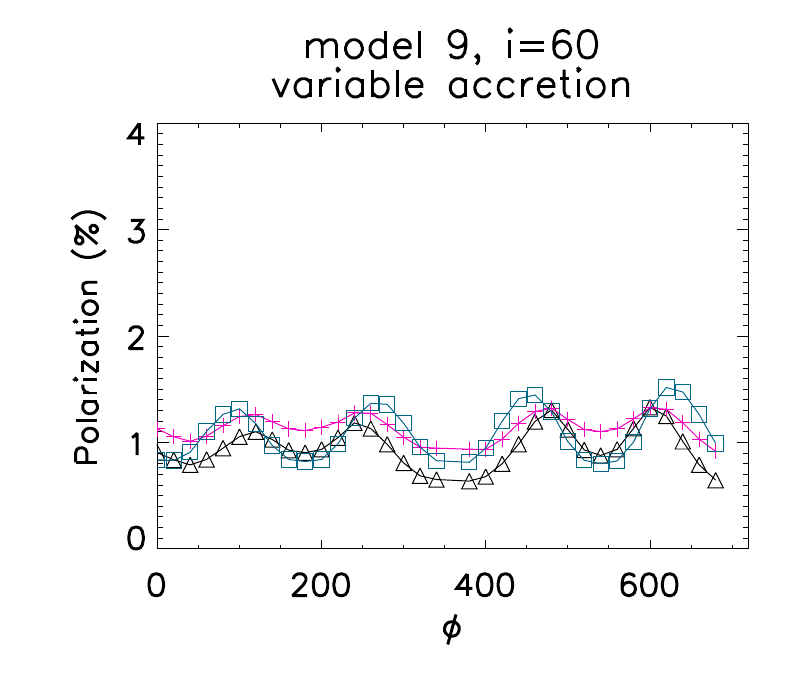}
\caption{\small 
Model polarization plots for the periodic category over two rotations, corresponding to the flux plots
of Figure~\ref{f:lc_period}.  
The symbols are black/triangle: V-band, pink/plus: I band, and blue/square: K band. 
}
\label{f:pol_period}
\end{center}
\end{figure*}

Our models that exhibit periodic behavior are shown in Figure \ref{f:lc_period} over two rotation periods. The first three light curves show 
the effects of changing the inclination angle ($i=90^\circ$ is edge-on viewing) on our initial model (Model 1). 
At small inclinations, the amplitude of variability at all wavelengths becomes smaller as the hotspots and inner disk wall are occulted less by the star compared to higher inclinations.  At $i= 60$\arcdeg, the light curve for Model 1 shows a secondary increase in magnitude 
where the spot on the lower hemisphere is viewed through the inner disk hole. At the lower inclinations of $40^\circ$ and $20^\circ$ the light curve is sinusoidal because only one hot spot is visible. 

At low inclinations ($i \la 40^\circ$) the optical and infrared variability is anti-correlated with optical dimming corresponding to infrared brightening. This is because at low inclinations the optical is dominated by the visible hotspot on the upper hemisphere. When this hot spot is out of view the optical dims but the infrared brightens as the warped inner edge of the disk may be seen and it is illuminated by the spot. 
At higher inclinations when the spot on the lower hemisphere is also in view then the infrared variability is correlated with the optical.

The next model, Model 2, demonstrates how lowering the hotspot flux (by decreasing the accretion rate)
decreases the amplitude of the $V$, $I$, and $J$ variations.  
Note that we follow equations 5 and 7 from \citet{whitney13} to set the spot parameters. The spot luminosity, $L_s$, is determined for a given 
disk accretion rate, $\dot{M}$, and inner disk truncation radius, $R_{\rm trunc}$, by 
$L_s = GM_\star\dot{M}[1/R_\ast-1/R_{\rm trunc}]$. For a given 
fractional spot coverage, $f_s$, the spot temperature is $T_s = T_\ast[1+L_s/(2L_\ast f_s)]^{1/4}$.

In Model 3, we increased the hotspot size from 0.7\% to 30\% of the stellar surface, 
which makes the spot temperature essentially the same as the stellar temperature 
resulting in little or no effect on the optical light curve.  However, the warm inner edge of the disk 
causes infrared variability as the disk rotates with the star. 
This is very similar to some published YSOVAR lightcurves 
which show no variability in the optical, but periodic variability in the infrared. An explanation for this behavior may be a 
complex accretion geometry \citep{romanova08, adams12}, producing many stellar hotspots, resulting in very small 
variability at short wavelengths (dominated by the star), 
but the warped accretion disk still produces infrared variability. Modeling a complex accretion geometry as a single hotspot can be justified because we can only observe the total light emitted by the surface of the star facing us (cannot resolve spots). Therefore, the important variables to model are spot covering fraction and temperature difference between the spot and the surface of the star. We find this method to reproduce consistent changes in magnitudes and overall trends compared to more complex hotspot geometry modeling (see Section \ref{highspot}).

Model 4 with a larger inner disk warp shows behavior similar to Model 3, except with more structure 
in the infrared variability, especially in the higher inclination model as the warp becomes more dominant
when the viewing angle is close to edge-on. 
Compared to Model 1 at the same viewing angle, the light curve of Model 5 demonstrates 
that a larger warp causes more variation at longer wavelengths.
Model 6 shows that with no disk warp, there is much less variation in the mid-infrared, 
with the near-infrared and optical variations about the same. 

Model 7 shows the case of a single hotspot (rather than two) and a warp. Relative to the two-spot model, the amplitude 
of the variability is larger at nearly all wavelengths and the lightcurves are less structured within their peaks and valleys. The 
infrared and optical light curves are clearly anti-correlated for such models with a single hotspot. 
In Model 8, we increased the inner disk radius, 
which created stronger variations in the IRAC data and weaker variations in the optical and near-infared light curves. 
Lastly, for Model 9, we increased the accretion rate steadily over two rotations, which created an overall 
upward trend in the brightness for all of the wavelengths. The trend is superposed on other azimuthal structure
that differs somewhat from Model 1 viewed at the same inclination due to the larger inner disk radius
that was also included in this model (as also seen in Model 8).


The variability described above is displayed quantitatively in Table \ref{periodicData}. We state changes in magnitudes (peak-to-peak) 
for all the periodic models at two wavelengths: optical ($V$ band) variability and mid-infrared variations 
(IRAC [3.6] band).  Full widths at half maximum (FWHM) are also reported for both optical and infrared, as well as whether the optical and IR variations are in-phase (correlated), out-of-phase (anti-correlated), or do not show any correlation (uncorrelated).
A range of different behaviors with wavelength are displayed in the models depending on the viewing angle and the projected geometry of the 
hotspots and disk structure. 
Some models have only infrared variability (Models 3 and 4) where the hot spots cover a large fraction of the star resulting in small optical variability. 
Other models display variability at all wavelengths with the amplitude of the 
variability being larger in the optical (Models 1, 2, 5, 6, 7), or a more complicated wavelength dependent variability (Models 8 and 9) that depends on 
the viewing angle towards the hotspots (that are responsible for the optical variability) and the warm inner edge of the disk warp 
(that dominates the infrared variability).
 
 Figure \ref{f:pol_period} shows the linear polarization light curves for the models described above. In general the polarization is around 
 0.5\% to 2\% which is typical for both observations \citep{perrin15} and models of dust scattering in YSO disks \citep{rob06,wh92}. The 
 amplitude of polarization variability in our models is typically less than about 1\%, again typical of models of rotationally modulated 
 polarization due to scattering of light from stellar hotspots \citep{wood96,stassun99}. The polarization degree is 
 lowest for low system inclinations such as Model 1 viewed at $i=20^\circ$, but the variability is greater than 2\% due to the asymmetry 
 of the hotspots illuminating different regions of the inner disk during the stellar rotation period. The polarization displays two maxima during 
 each rotation period which arise when the spots are on the limb of the star (twice per rotation) and the light from the hot spots is scattered 
 into our line of sight at angles close to $90^\circ$ (where polarization due to scattering is maximum). The single maximum in the intensity 
 and double maximum in polarization light curves are clear signatures of variability due to hotspots and if observed would lend further 
 support for the models we have presented.

In summary, the models presented in this section are intended to correspond to the YSOVAR lightcurves that
exhibit multiwavelength variability in a periodic or quasi-periodic fashion.  Changing model parameters
from the initial model (Model 1) had the effect of changing both the relative flux variation at the different wavelengths,
and the lightcurve shape.  The lightcurves output from the models generally retained the periodic nature 
imposed by the dominant dynamical effect of stellar/magnetosphere rotation.  
However, the addition of a variable mass accretion rate occurring on timescales comparable to 
the rotation period (Model 9) rendered the output lightcurves only quasi-periodic
rather than strictly periodic.

\begin{deluxetable*}{c c c c c c l}
\tablecolumns{6}
\centering
\tablewidth{0pt}
\tablecaption{Statistics for the Periodic Models} 

\tablehead{\colhead{ } &\colhead{Inclination} & \colhead{$\Delta$mag\tablenotemark{1}} & \colhead{$\Delta$mag} & \colhead{FWHM\tablenotemark{2}} & \colhead{FWHM} & \colhead{Correlation}
\\
\colhead{} & \colhead{Angle} & \colhead{(V-Band)} & \colhead{(IRAC[3.6])} & \colhead{(V-Band)} & \colhead{(IRAC[3.6])} & \colhead{(IR vs. Optical)} }

\startdata
                 
Model 1& $20\arcdeg$& 0.35 & 0.05 & 50\% & 31\% & anti-correlated\\ 
Model 1 & $40\arcdeg$ & 0.65 & 0.12 & 51\% & 39\%& anti-correlated \\
Model 1 & $60\arcdeg$ & 0.65 & 0.17 & 57\% & 37\%& correlated\\
Model 2 & $60\arcdeg$ &0.3 & 0.12 & 60\% & 14\%& correlated \\
Model 3 & $60\arcdeg$ &0.02 & 0.08 & 23\% & 14\%& uncorrelated  \\
Model 4 & $60\arcdeg$ & 0.05 & 0.13 & 78\% & 30\%& uncorrelated\\ 
Model 4 & $75\arcdeg$ & 0.075 & 0.9 & 31\% & 73\%& uncorrelated \\
Model 5 & $60\arcdeg$ & 0.65 & 0.15 & 59\% & 26\%& correlated \\
Model 6 & $60\arcdeg$ & 0.67 & 0.05 & 59\% & 25\%& correlated \\
Model 7 & $60\arcdeg$ & 1.1 & 0.4 & 42\% & 43\%& anti-correlated \\
Model 8 & $60\arcdeg$ & 0.67 & 0.25 & 60\% & 19\%& correlated \\
Model 9\tablenotemark{3} & $60\arcdeg$ & - & - & - & -& uncorrelated \\ [1ex]
\enddata

\tablenotetext{1}{Magnitude change for largest peak-to-peak variation}
\tablenotetext{2}{FWHM measurement is for the largest peak-to-peak periodic dip}
\tablenotetext{3}{No magnitude changes given for this model because accretion rate and overall brightness steadily increases over two rotation periods}
\label{periodicData}
\end{deluxetable*}

\subsection{Periodic Dippers} 
Periodic dippers show a relatively steady flux followed by regularly spaced dimming (dipping) events that last from around a day to a week. 
It has been suggested that the longer timescale events are caused by a warped disk passing across the line of sight 
and thus obscuring the star \citep{bertout00,bouvier03, mcginnis2015}. 
The multi-wavelength observations of \citet{morales11}, \citet{cody14}, and \citet{stauffer15}, 
suggest that in general the dipper light curves 
exhibit greater variability in the optical than in the infrared. \citet{cody14} note that in the joint Spitzer/CoRoT 
sample, 35 sources displayed optical dips compared to seven that also displayed infrared dips and only two that had dips 
only in the infrared.  This is understandable if the dips are due to extinction from dust in the line of sight, which
would produce the greatest variations at optical wavelengths according to typical extinction laws.

\begin{figure*}[h,t!]
\begin{center}
\includegraphics[scale=0.7]{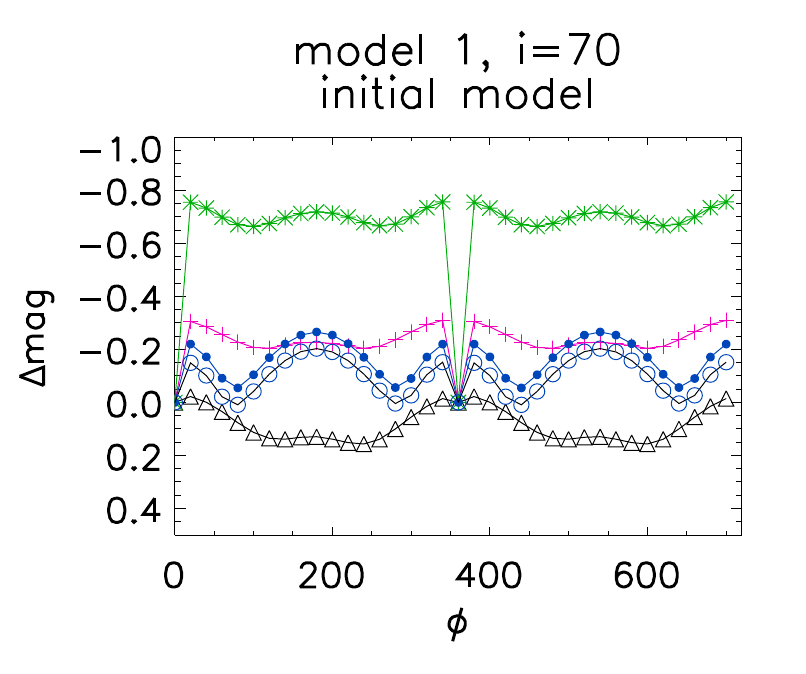}
\includegraphics[scale=0.7]{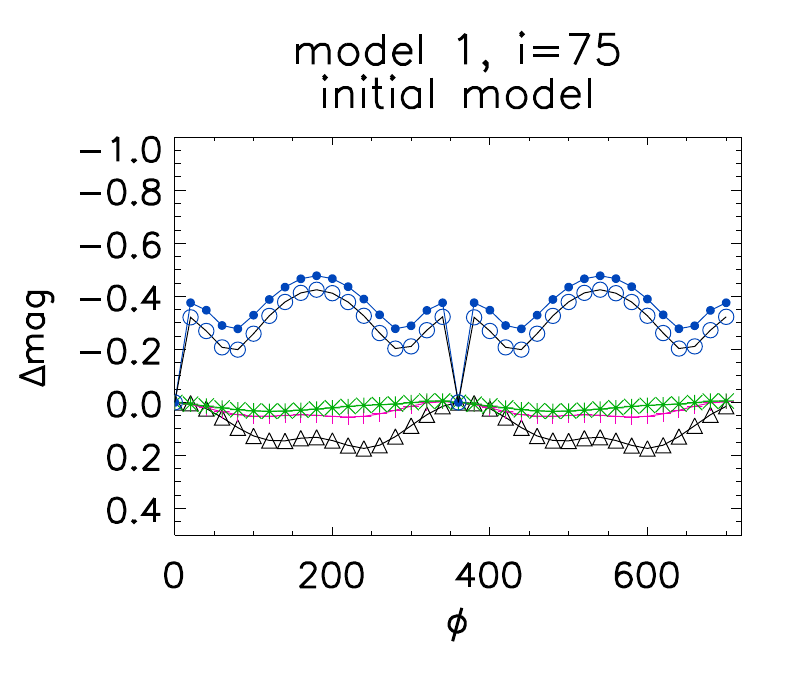}
\includegraphics[scale=0.7]{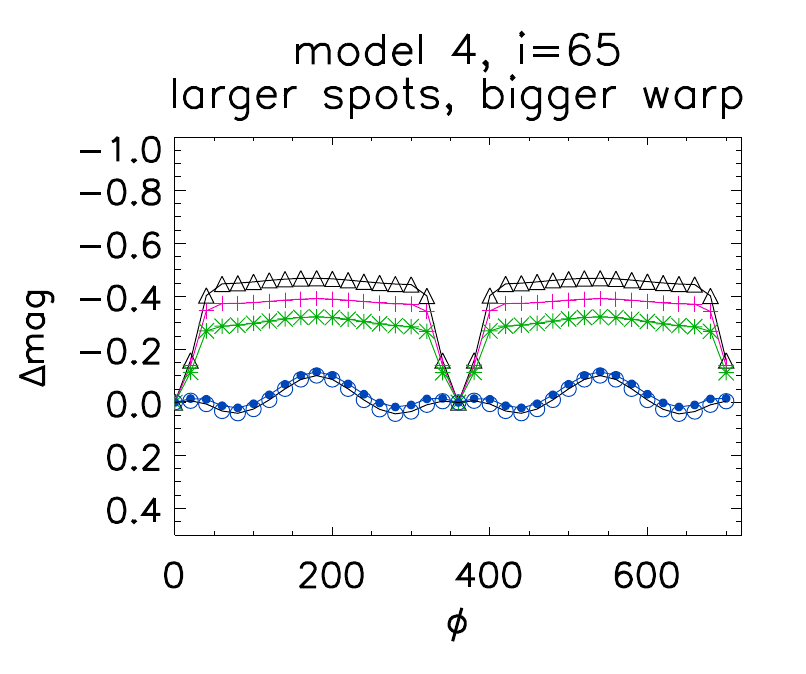}
\includegraphics[scale=0.7]{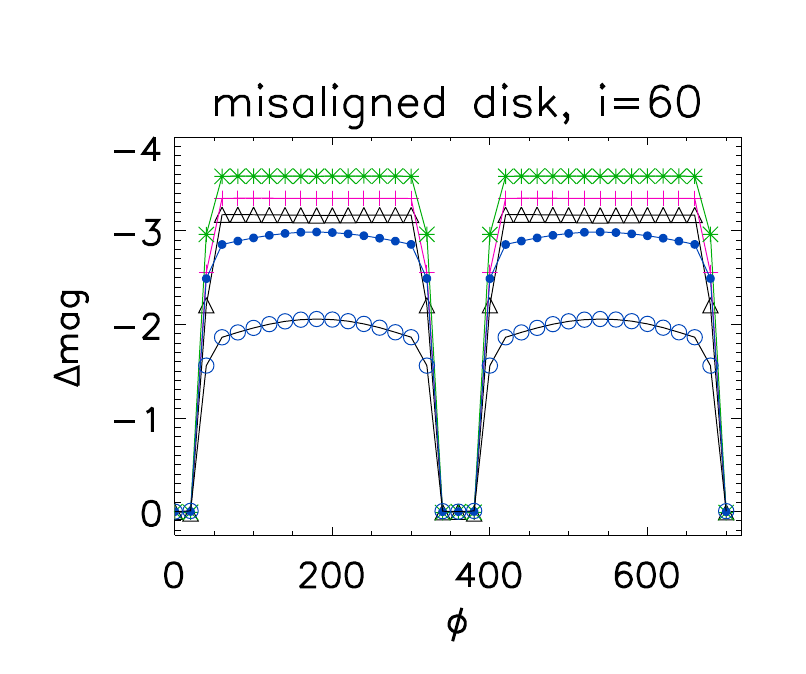}
\caption{\small
Model light curves that can reproduce features of objects in the dipper category; 
see Table 1 for model details, with the misaligned inner disk model taken from \citet{whitney13}. 
Symbols are the same as in Figure \ref{f:lc_period}. 
}
\label{f:lc_dip}
\end{center}
\end{figure*}

\begin{figure*}[h,t!]
\begin{center}
\includegraphics[scale=0.7]{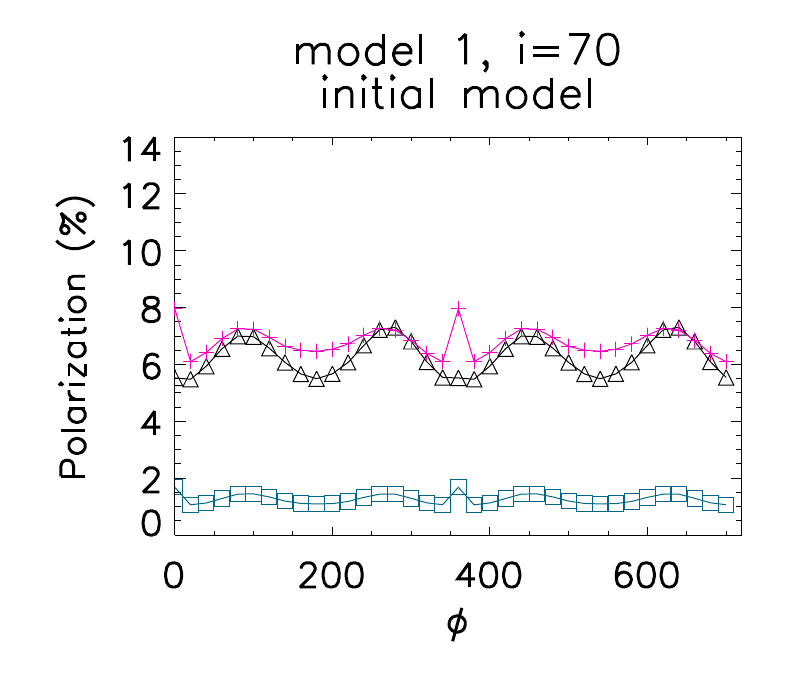}
\includegraphics[scale=0.7]{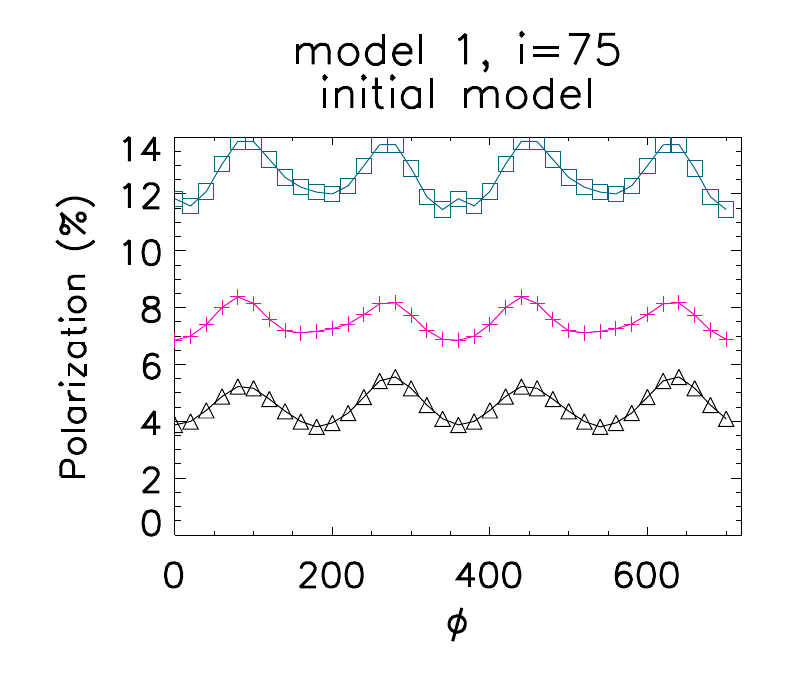}
\includegraphics[scale=0.7]{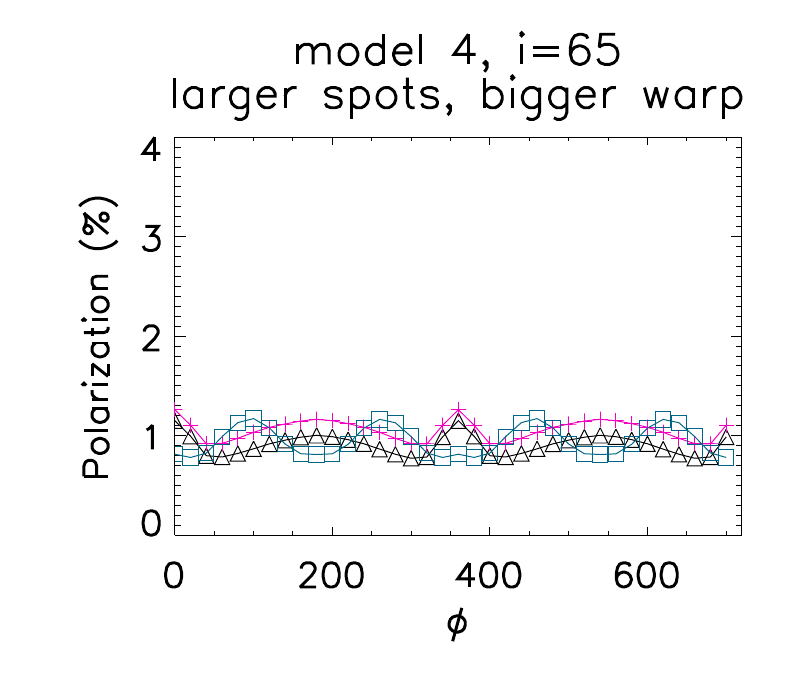}
\includegraphics[scale=0.7]{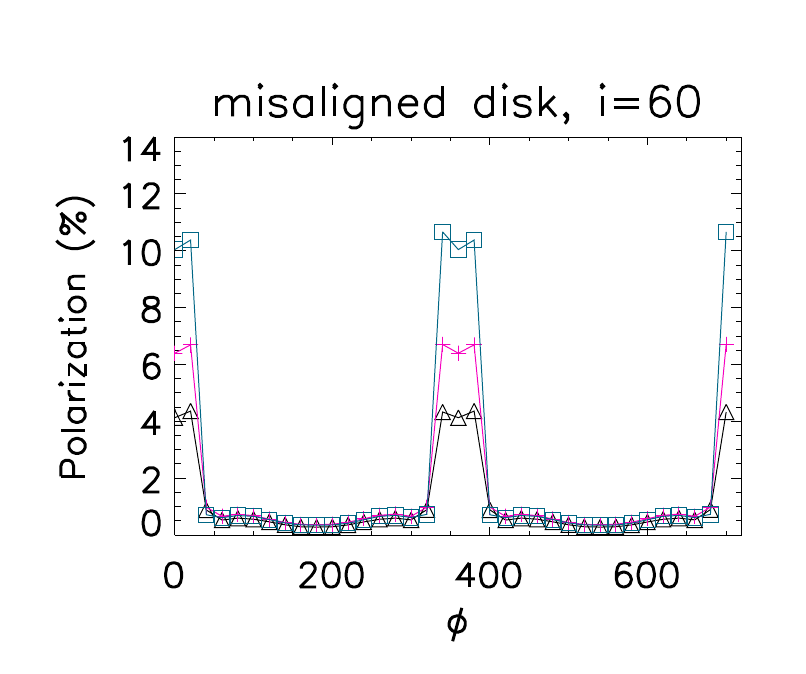}
\caption{\small
Model polarization plots over two rotations, corresponding to the flux plots
of Figure~\ref{f:lc_dip} with polarization of the misaligned disk from \citet{whitney13}. 
Symbols are the same as in \ref{f:pol_period}. 
}
\label{f:pol_dip}
\end{center}
\end{figure*}

In order for dips to occur in our model, the viewing angle needs to be close to edge-on so that over the
azimuthal range where the disk is not warped, it does not obscure the star, 
and when the warp passes through the line of sight as the star/disk system rotates, it obscures the star.
We can achieve dips in both the optical and mid-infrared data by changing the inclination of the models. In Figure \ref{f:lc_dip}, 
Model 1 at $i=70^\circ$ exhibits a $V$-band brightness level that is heavily extincted throughout the 
entire light curve because of the dusty disk, with the $I$ band exhibiting a similar shape, but less faint. 
The $J$ band has the greatest dip in magnitude because at this wavelength the radiation is emitted mostly by the star and is not extincted 
until the warp passes in front of it.  The infrared
bands  show only a small dip from the extinction in the warp and otherwise show the periodic pattern based on the
projected area of the visible disk inner wall.    
In Model 1 shown at a viewing angle of 75\arcdeg, the optical light is 
completely extincted but there is a large dip at IRAC wavelengths because when the warp passes into the line of sight 
it also obscures the back of the disk which emits in the mid-infrared bands.    
When Model 4 is viewed at $i=65^\circ$ none of the bands are extincted until the warp passes through the line of sight,
with the $V$ band having the greatest variability since it is dominated by (obscured) emission from the star.
We include for comparison a model from \citet{whitney13} which also displays periodic dips at all wavelengths. In this model, the inner disk is 
misaligned by $30\arcdeg$ with respect to the outer disk. Because of this misalignment, the inner disk blocks the star from view as it rotates through certain azimuths.  

The corresponding polarization variations are illustrated in Figure \ref{f:pol_dip} and, for the most part, show increased percentages of polarization for the lightcurve models exhibiting the dipper behavior.  This is because of the large viewing angles above $70\arcdeg$ that sample lines of sight through the disk. With the misaligned disk we are also looking directly through the inner disk at certain times during the rotation period. 

The models presented in this section are intended to match the ``dipper" category of YSOVAR variables where the
fading events require variable obscuration by an inner disk warp or a misaligned disk.  In addition to the
models illustrated here with variations at all wavelengths, among our full model set are cases 
where there are dips in the visible but not in the infrared, others where only the IRAC wavelengths 
exhibit extinction events, and still others where both the IRAC and visible bands exhibit periodic extinction.  
\citet{cody14} and \citet{morales11} found all three cases in their datasets.

\subsection{Irregular Dippers}
 
Many of the YSOVAR light curves show states of modest photometric variation followed by sharp drops in brightness, 
but do not exhibit the essentially periodic dipper-like variations discussed in the previous section.
Instead the dips are quite irregular.  Some of these ``aperiodic dipper" or ``stochastic plus dipper" light curves 
may be caused by extinction events similar to those in the periodic dipper category.  The difference is that
instead of having the extinction events at regular time intervals and similar magnitude changes, they are more 
stochastic and unpredictable, with significant stochastic behavior in the light curve outside of the dip as well.  
Figure 11 in \citet{cody14} shows examples of this type of light curve. 
In order to explain the irregular but asymmetrically fading variations,
 we assume an unstable accretion model such as that proposed by \citet{romanova08}. 
Because of the many different streams of infalling material, a highly variable light curve results, which we recreate 
in our models with a parameter that changes the fractal clumping of the accretion disk 
\citep[][Section 3.7]{whitney13}. 

\begin{figure*}[h,t!]
\begin{center}
\includegraphics[scale=0.7]{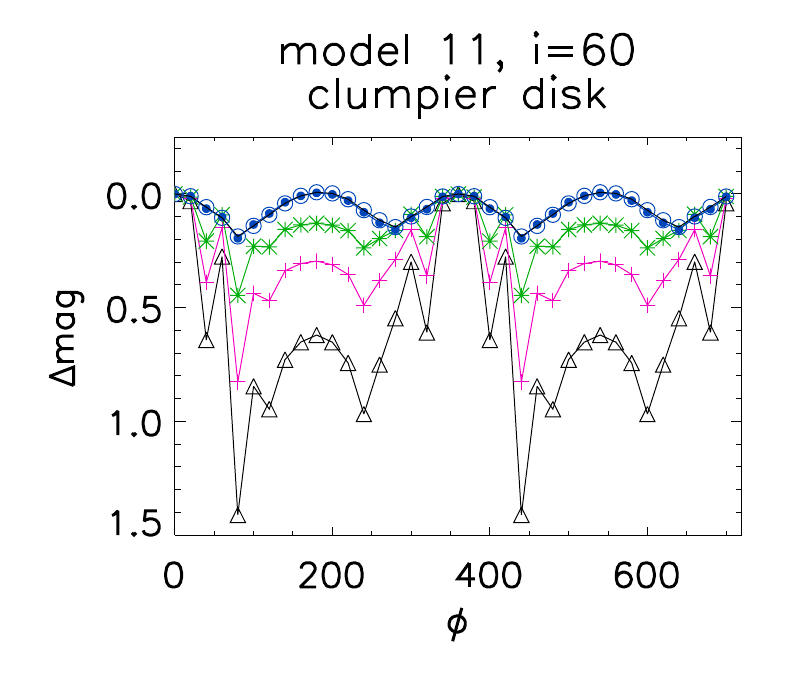}
\includegraphics[scale=0.7]{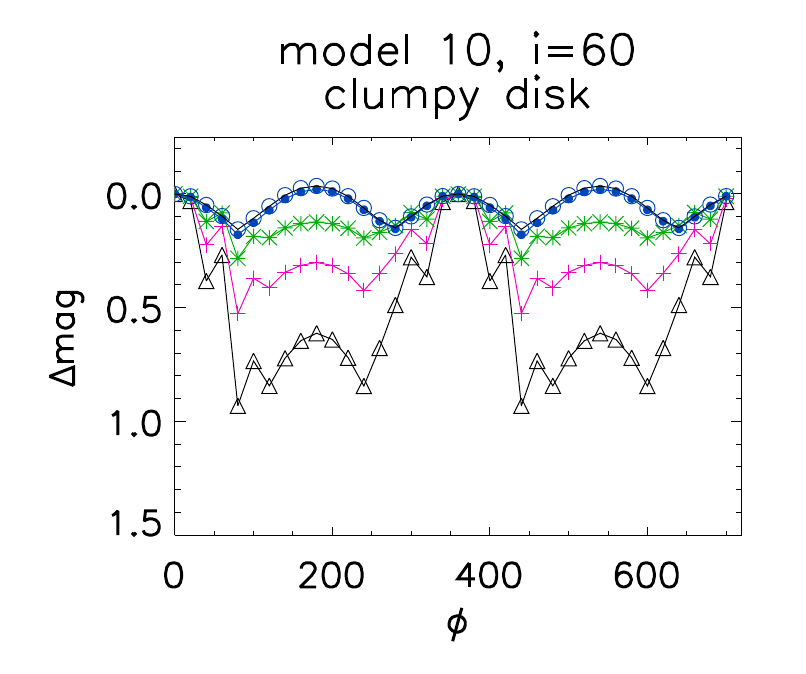}
\includegraphics[scale=0.7]{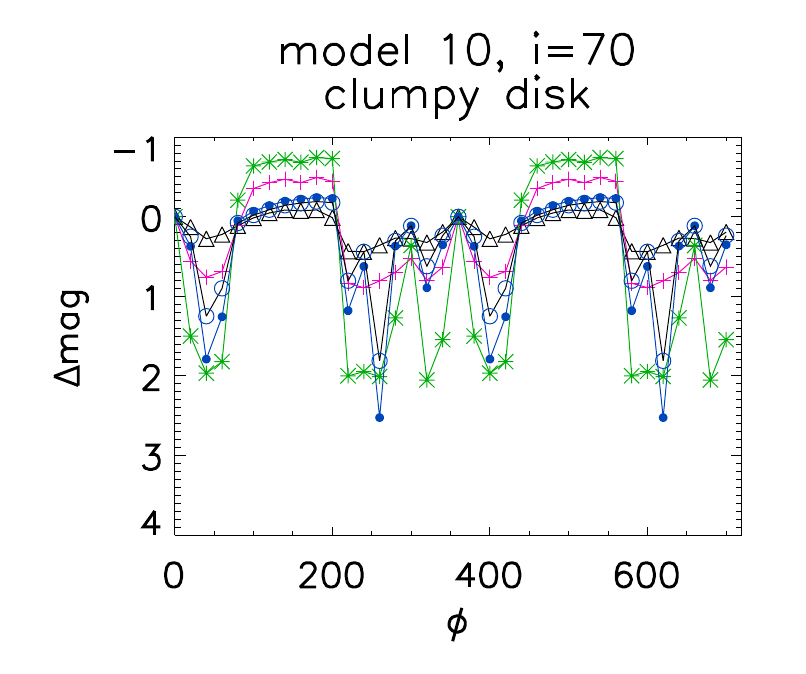}
\includegraphics[scale=0.7]{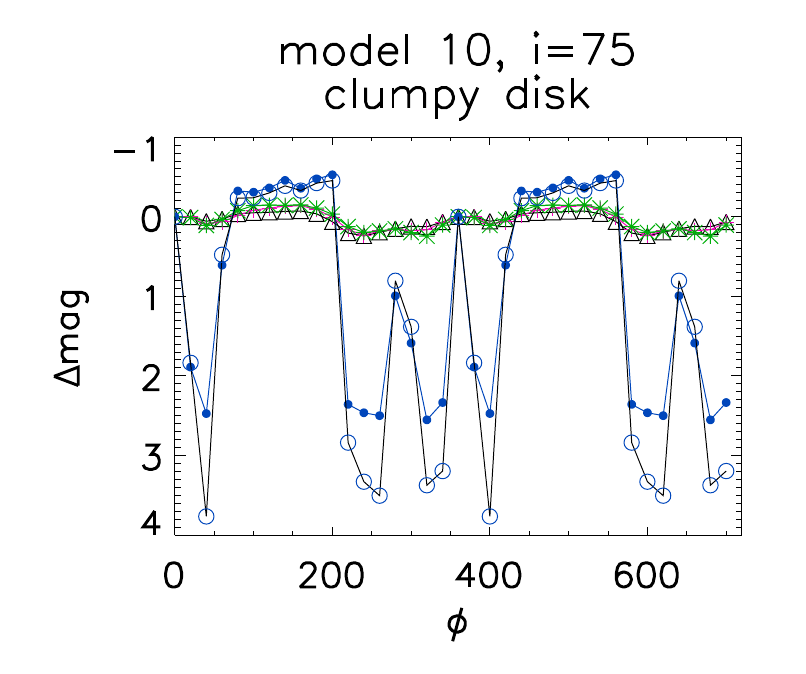}
\caption{\small
Model light curves for the irregular dipper category over two periods.
Symbols are the same as in Figure \ref{f:lc_period}}
\label{f:lc_wild}
\end{center}
\end{figure*}

\begin{figure*}[h,t!]
\begin{center}
\includegraphics[scale=0.7]{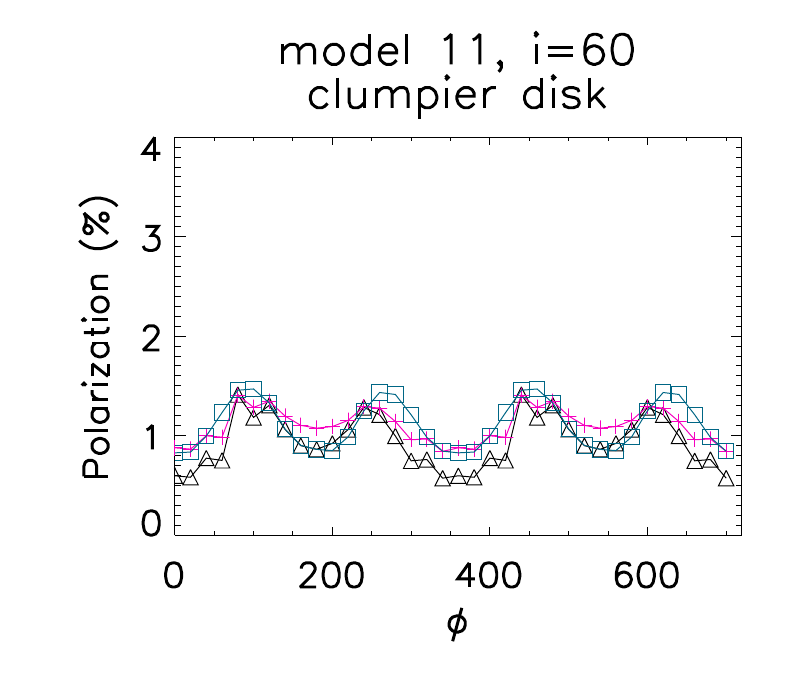}
\includegraphics[scale=0.7]{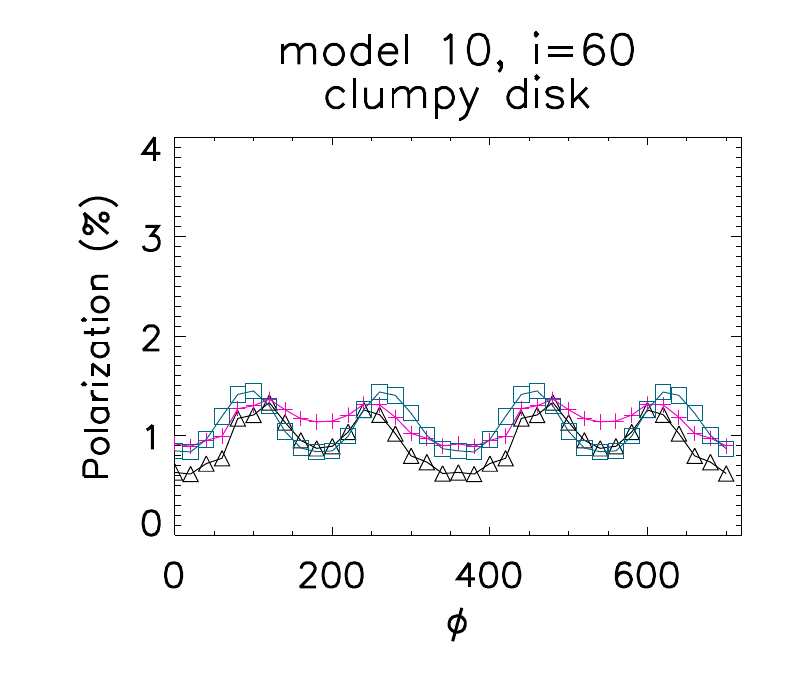}
\includegraphics[scale=0.7]{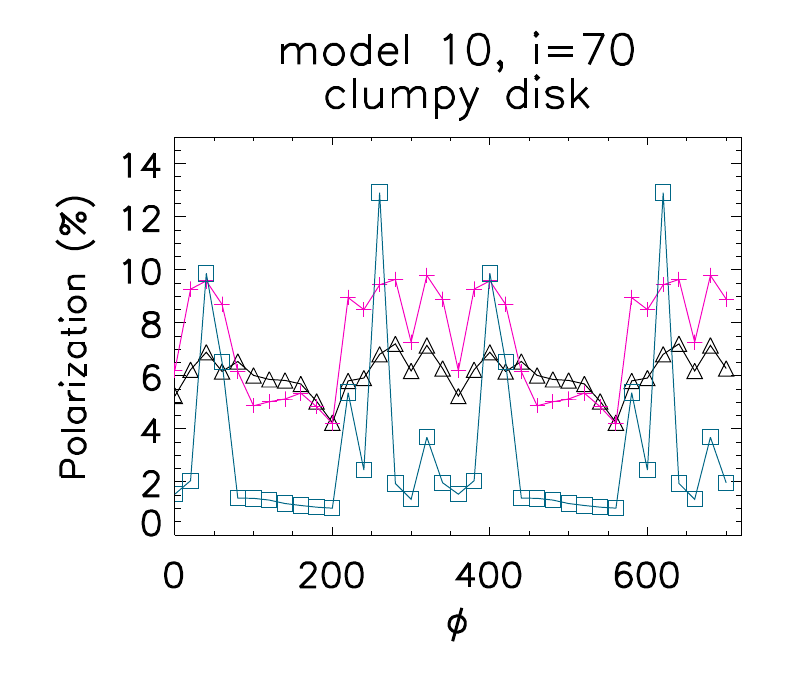}
\includegraphics[scale=0.7]{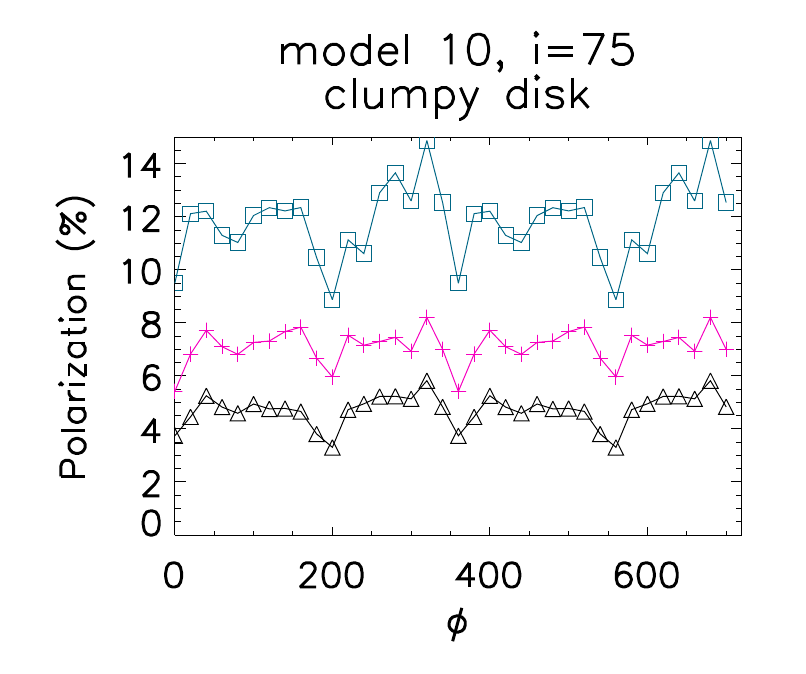}
\caption{\small
Model polarization plots for the irregular dipper category over two periods.
Symbols are the same as in Figure \ref{f:pol_period}}
\label{f:pol_wild}
\end{center}
\end{figure*}

Our models that demonstrate aperiodic dipper type variations are shown in Figure \ref{f:lc_wild}, and are
presented as Models 10 and 11 in Table 1.  The polarization plots for the same models are shown in Figure \ref{f:pol_wild}.  
At viewing angles $i\la 60^\circ$, we are not looking directly through the disk 
so the effects of the clumped disk are not very obvious, except in the visible band.
As the inclination increases to 70\arcdeg, the effects from the disk clumping become much more pronounced
and all of the bands show irregular flux variations. For  $i=75^\circ$  
the visible bands are mostly extincted and thus show less variation while the infrared bands are not subject to extinction
and show more variations.  
The differences between Model 10 and Model 11 are a result of changing the fraction of material 
that is clumped. When the clumping is increased from Model 10 to Model 11, the variations become larger.

\subsection{High Latitude Hotspots}
\label{highspot}
\begin{figure*}[h,t!]
\begin{center}
\includegraphics[scale=0.6]{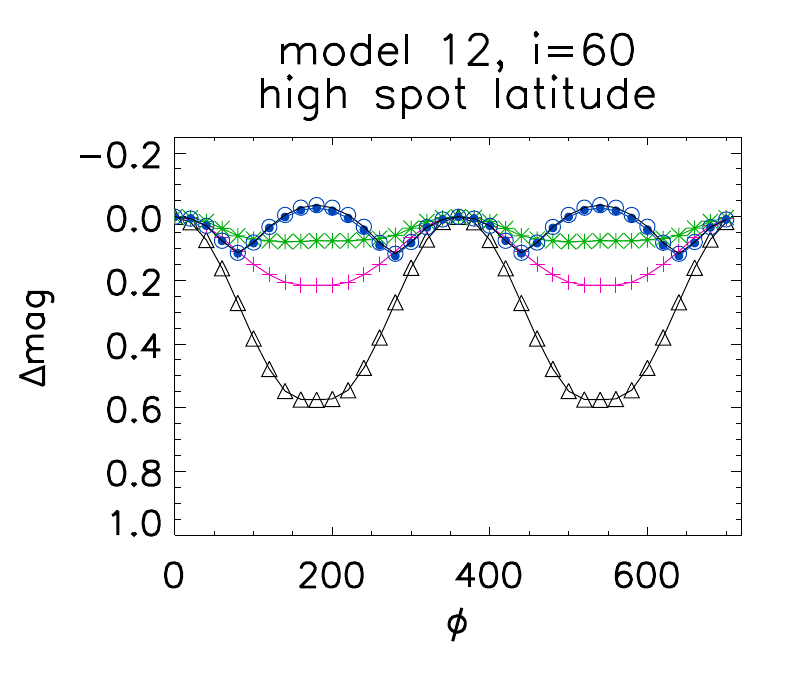}
\includegraphics[scale=0.6]{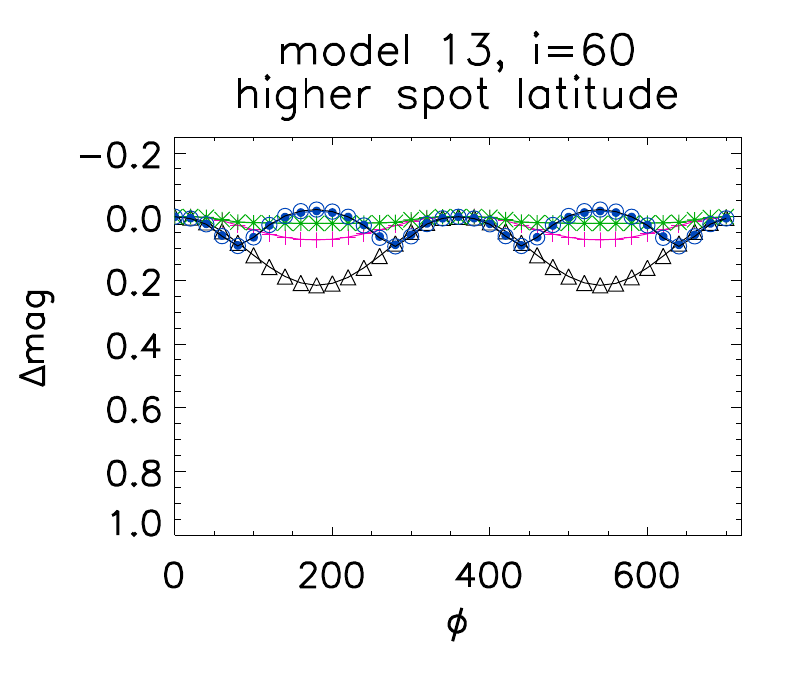}
\includegraphics[scale=0.6]{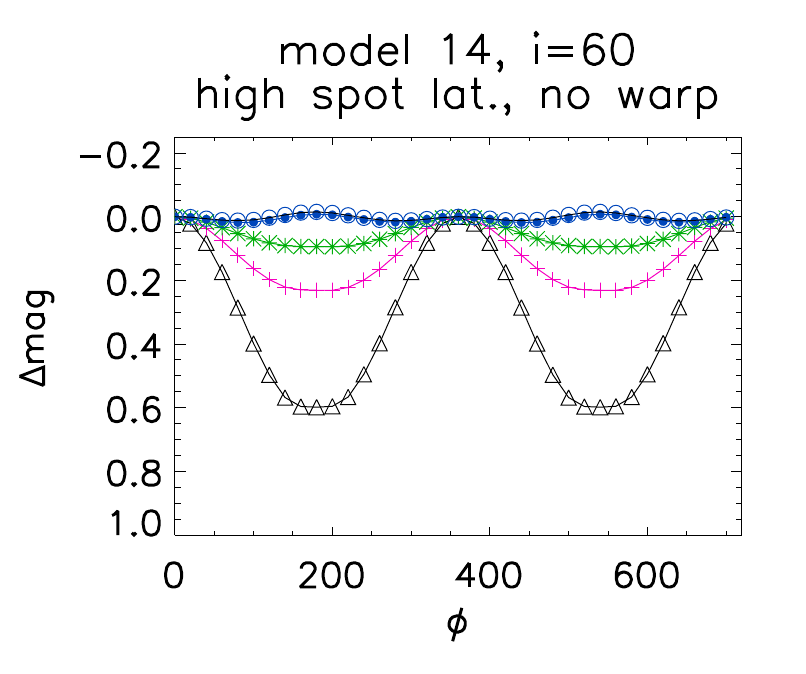}
\includegraphics[scale=0.6]{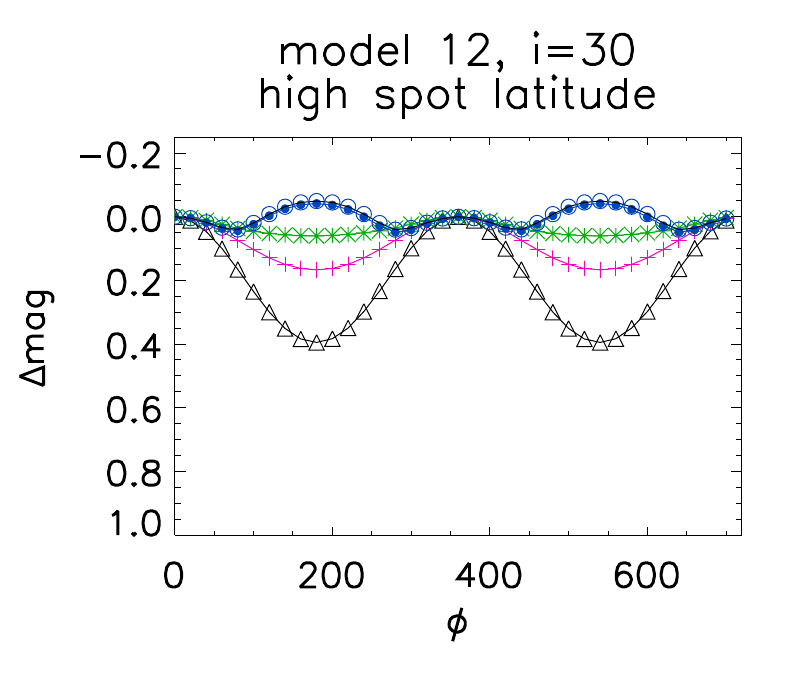}
\includegraphics[scale=0.6]{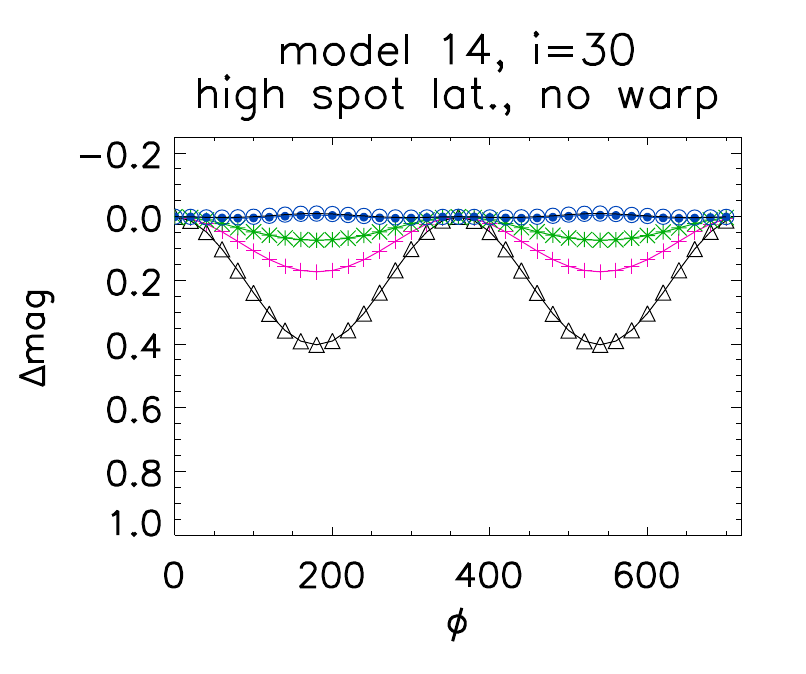}
\includegraphics[scale=0.6]{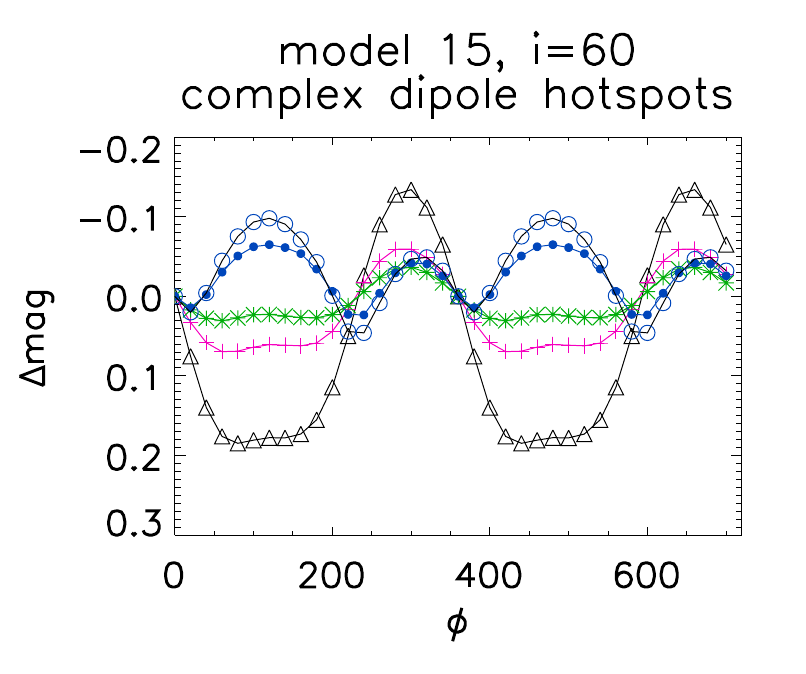}
\includegraphics[scale=0.6]{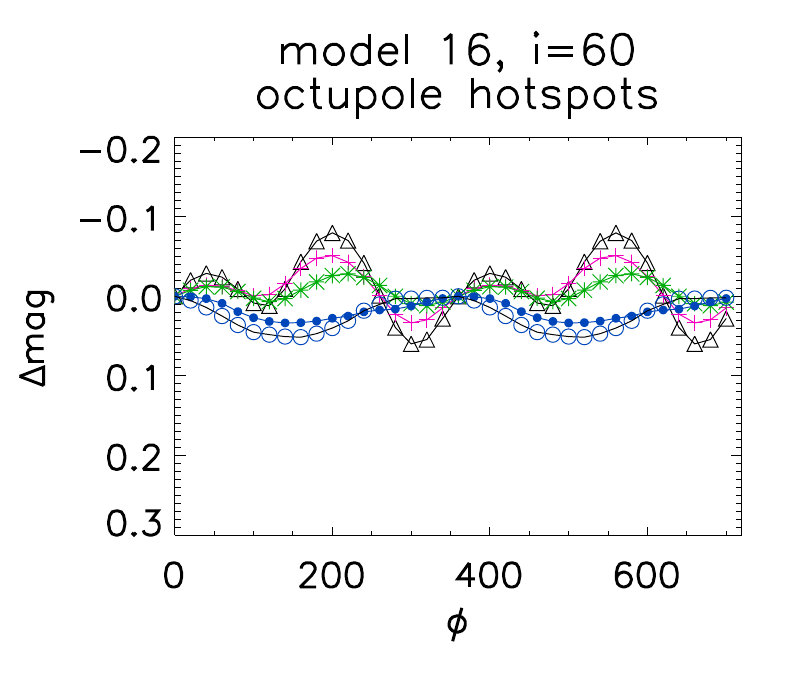}
\includegraphics[scale=0.6]{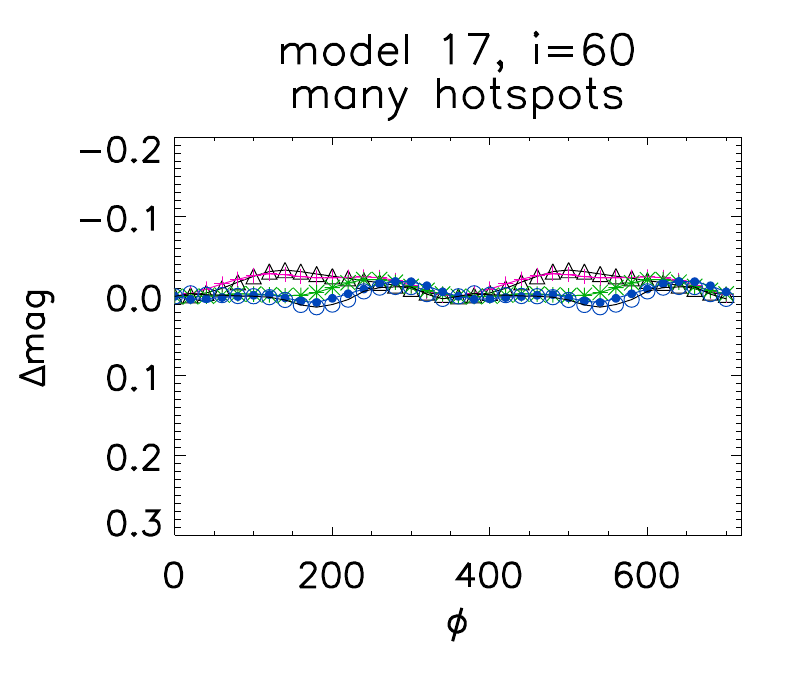}
\caption{\small
Model light curves over two rotation periods for the models with high spot latitudes (see Section 3.4 for detailed description).
Symbols are the same as in Figure \ref{f:lc_period}. 
}
\label{f:lc_highspot}
\end{center}
\end{figure*}

\begin{figure*}[h,t!]
\begin{center}
\includegraphics[scale=0.6]{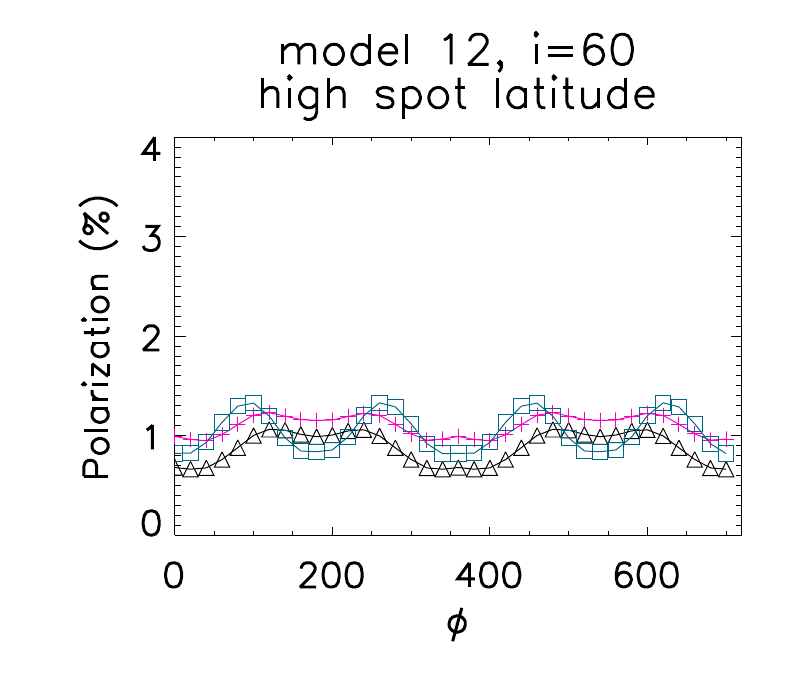}
\includegraphics[scale=0.6]{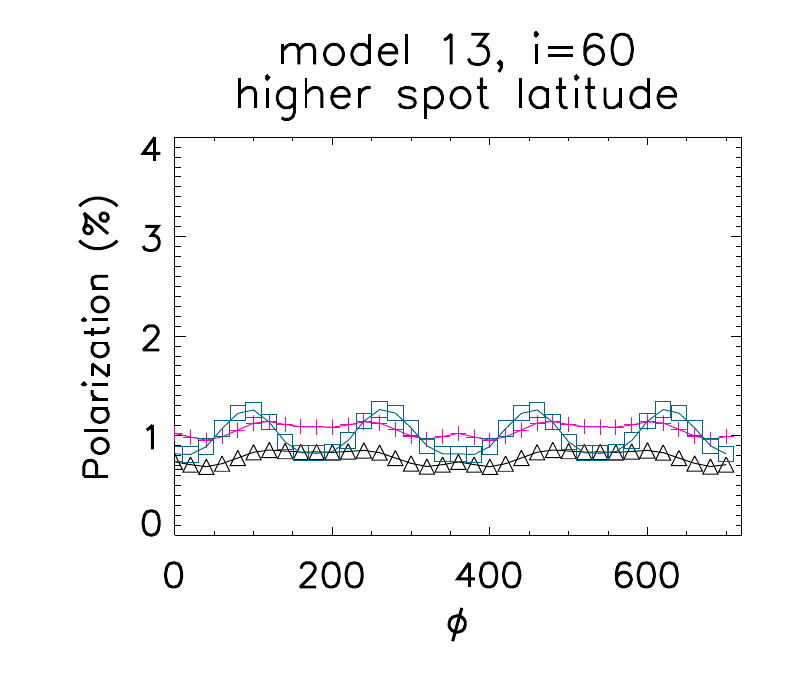}
\includegraphics[scale=0.6]{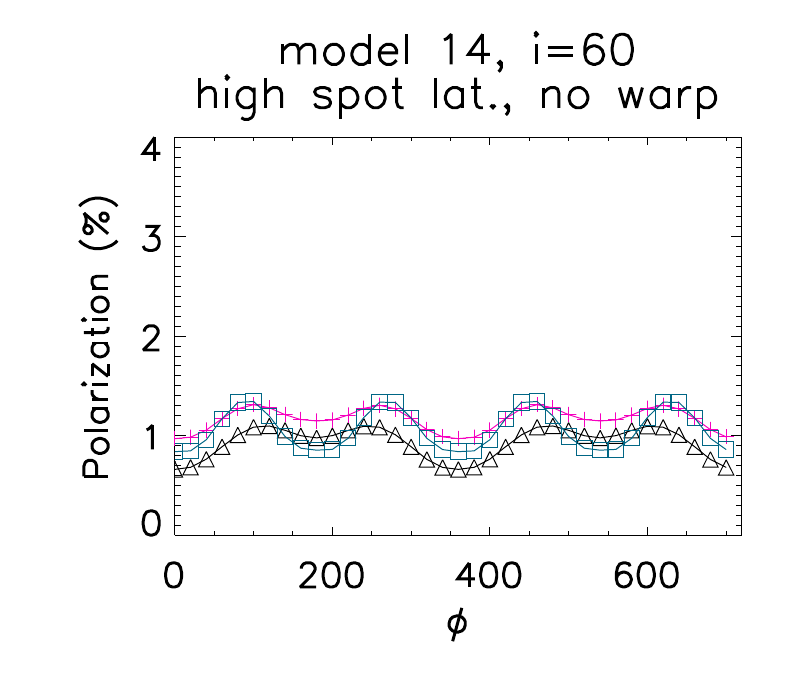}
\includegraphics[scale=0.6]{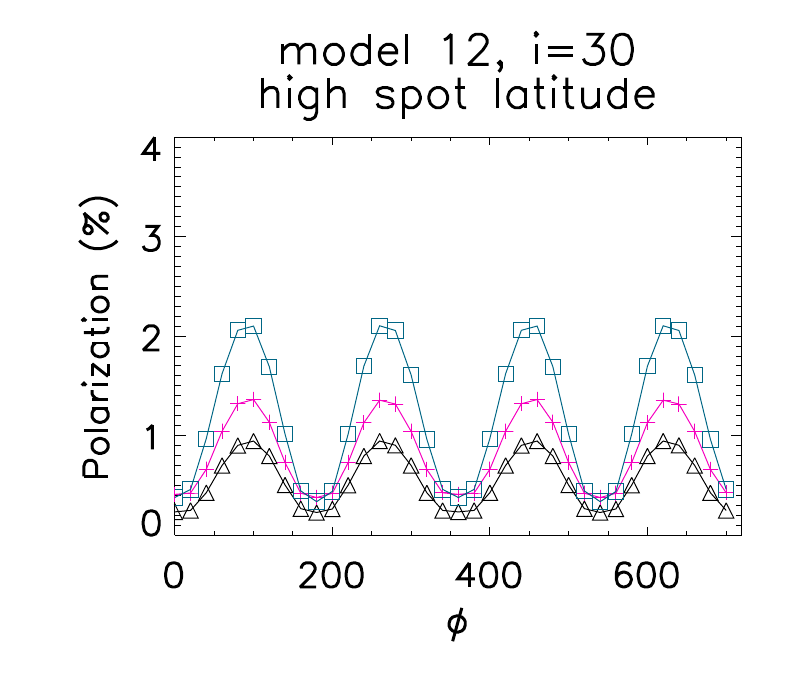}
\includegraphics[scale=0.6]{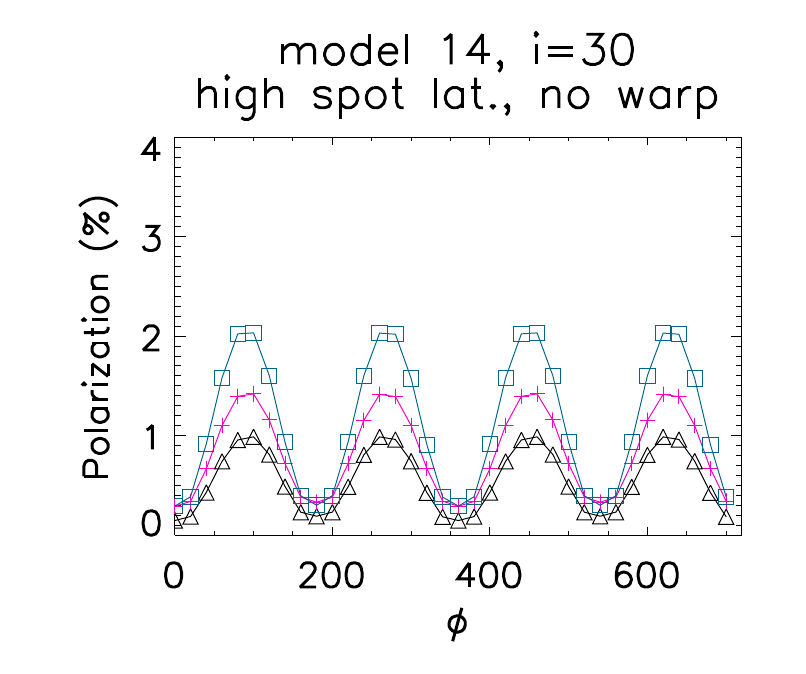}
\includegraphics[scale=0.6]{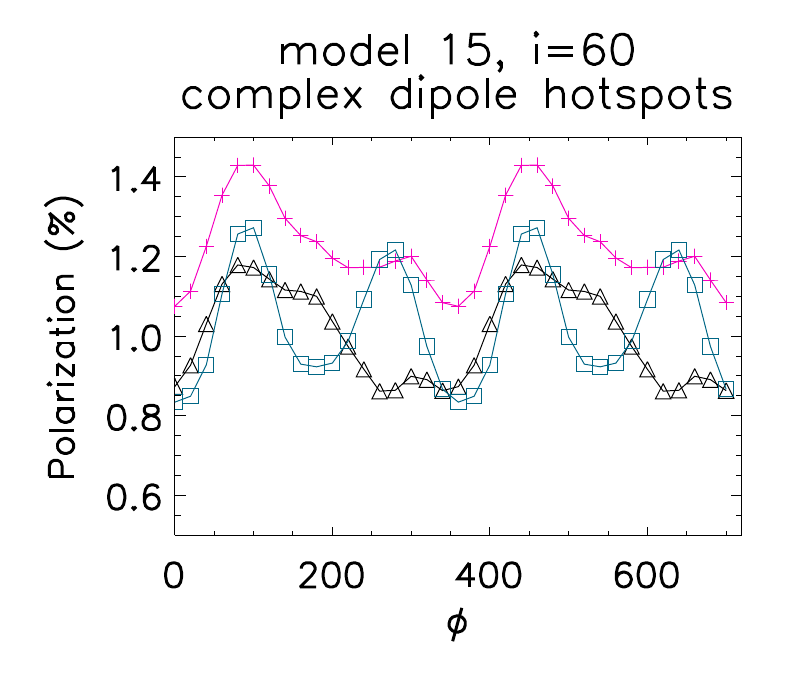}
\includegraphics[scale=0.6]{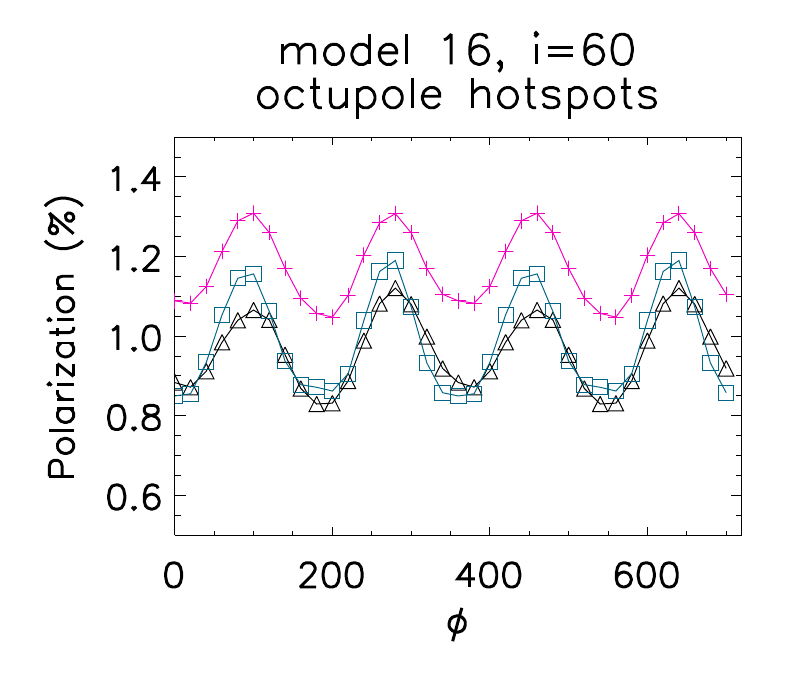}
\includegraphics[scale=0.6]{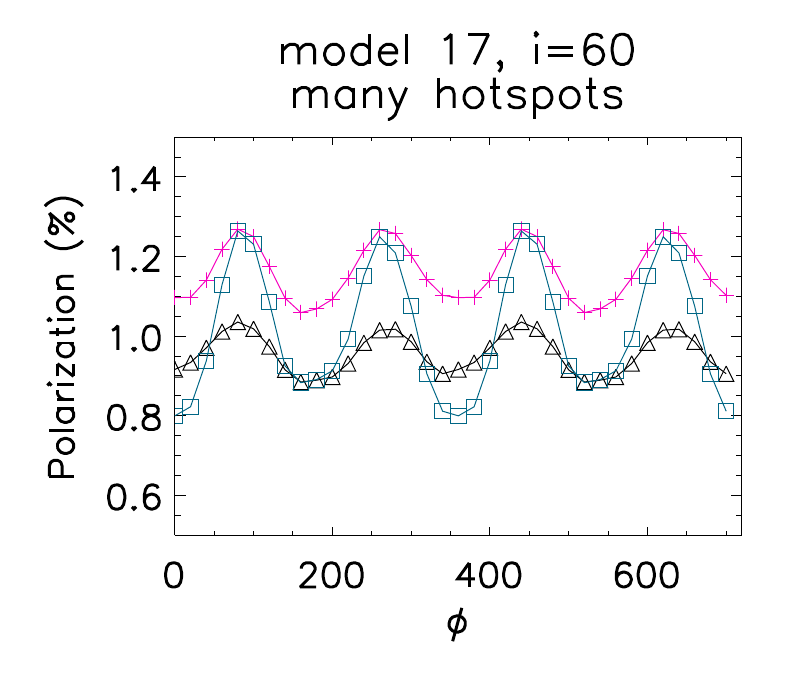}
\caption{\small
Model polarization plots over two rotation periods for the high spot latitude models.
Symbols are the same as in Figure \ref{f:pol_period}. 
}
\label{f:pol_highspot}
\end{center}
\end{figure*}

We created a subset of models that were motivated by the fact YSOVAR observations indicate mostly in-phase or uncorrelated behavior between the optical and infrared lightcurves (see Section~4), whereas some of our periodic models at low inclinations exhibit anti-correlated behavior of the IRAC and optical bands. Models 12 and 13 both employ the same geometry as model 1, except the hotspots are situated at higher latitudes (60\arcdeg or 80\arcdeg). Model 14 does not include a warp in the accretion disk (similar to model 6) and again has hotspots at the high latitude of 60\arcdeg. The light curves for these models are shown in Figure \ref{f:lc_highspot}, and the polarization plots in Figure \ref{f:pol_highspot}. Models 12 and 13 both exhibit periodic trends, and still demonstrate the anti-correlated behavior between the IRAC and optical bands. Model 14 has approximately zero variability in the infrared due to the fact the disk is not warped and most of the infrared light comes from the disk.

\begin{figure*}[h,t!]
\begin{center}
\includegraphics[scale=0.4]{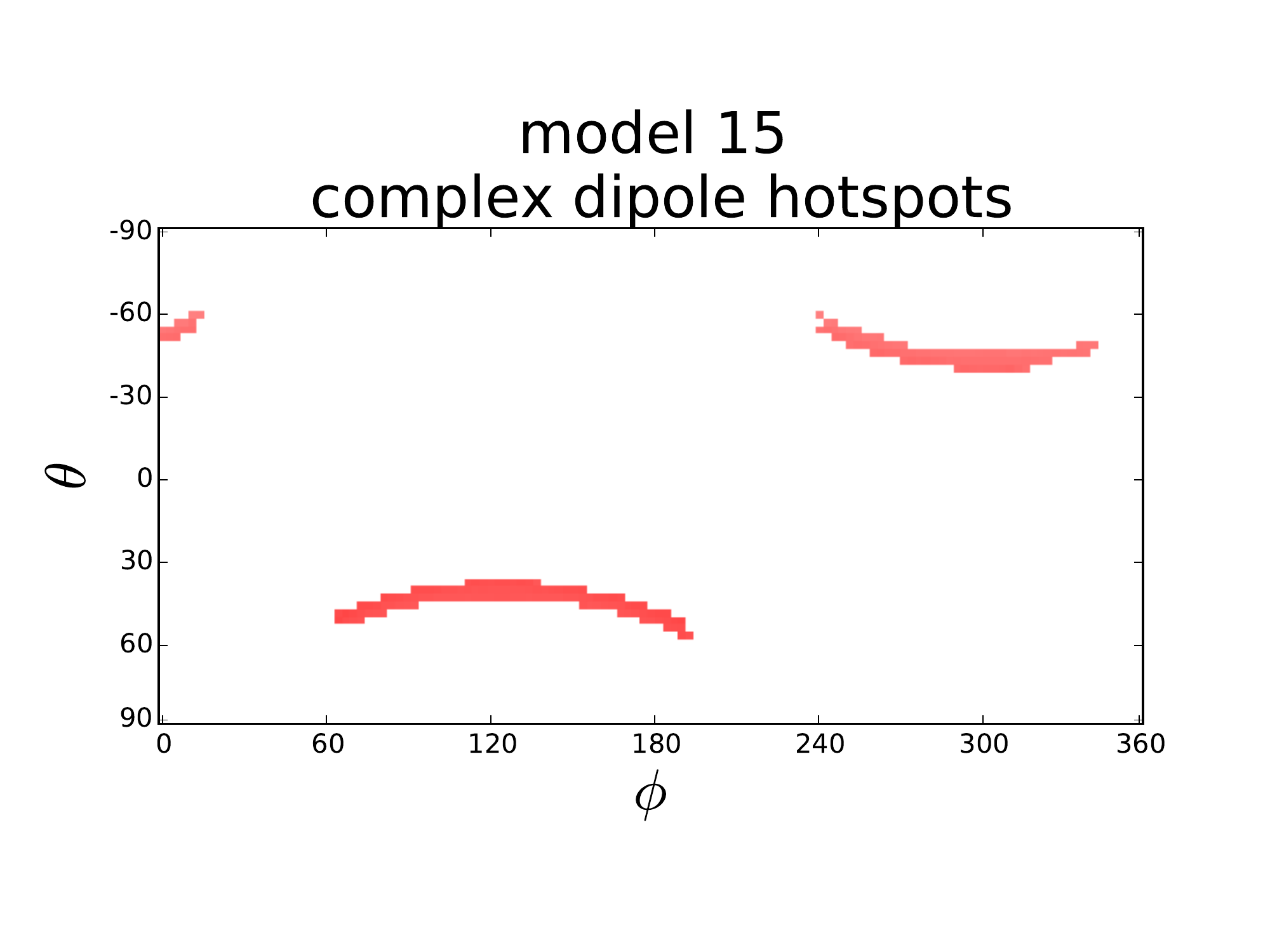}
\includegraphics[scale=0.4]{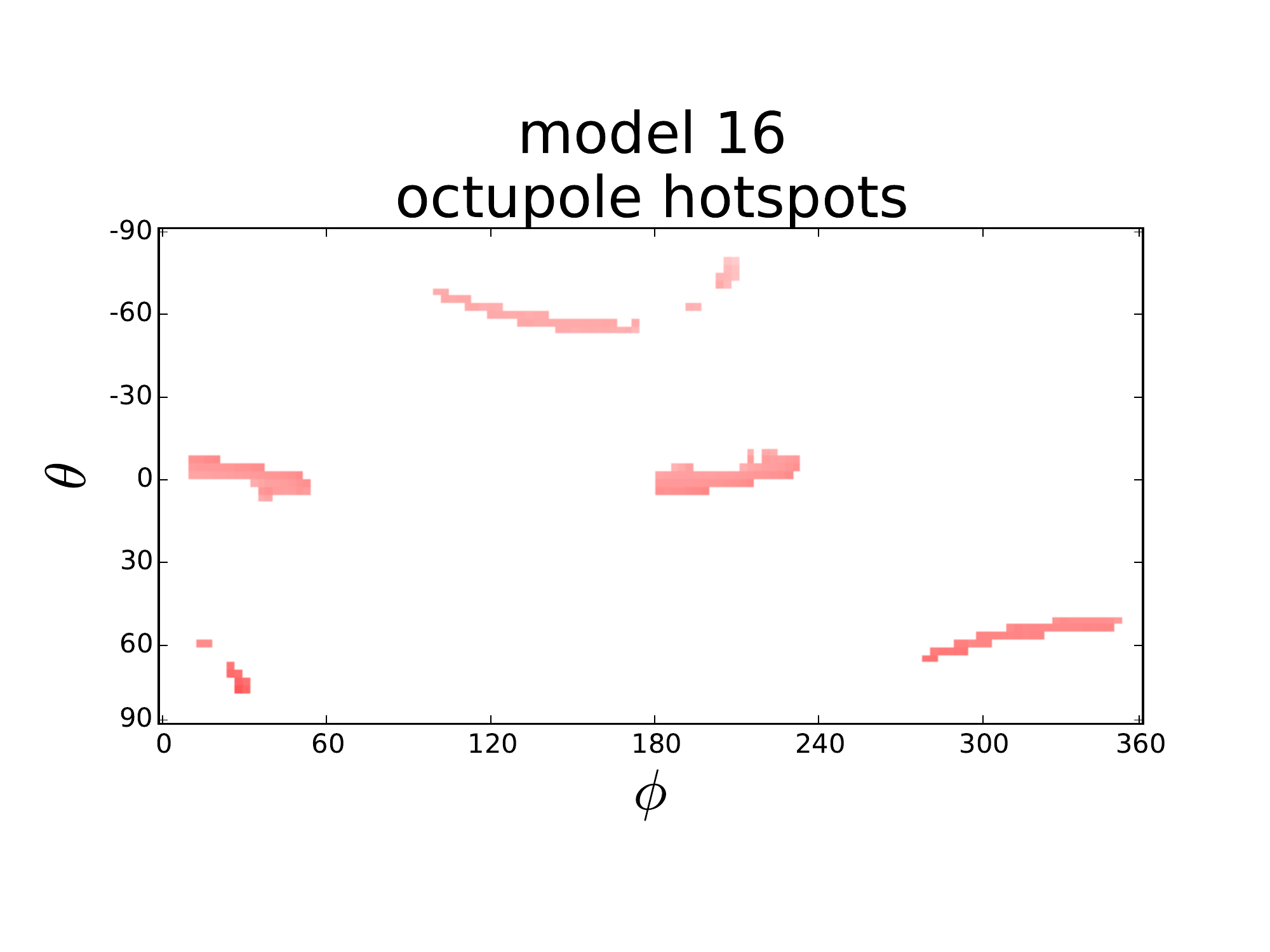}
\includegraphics[scale=0.4]{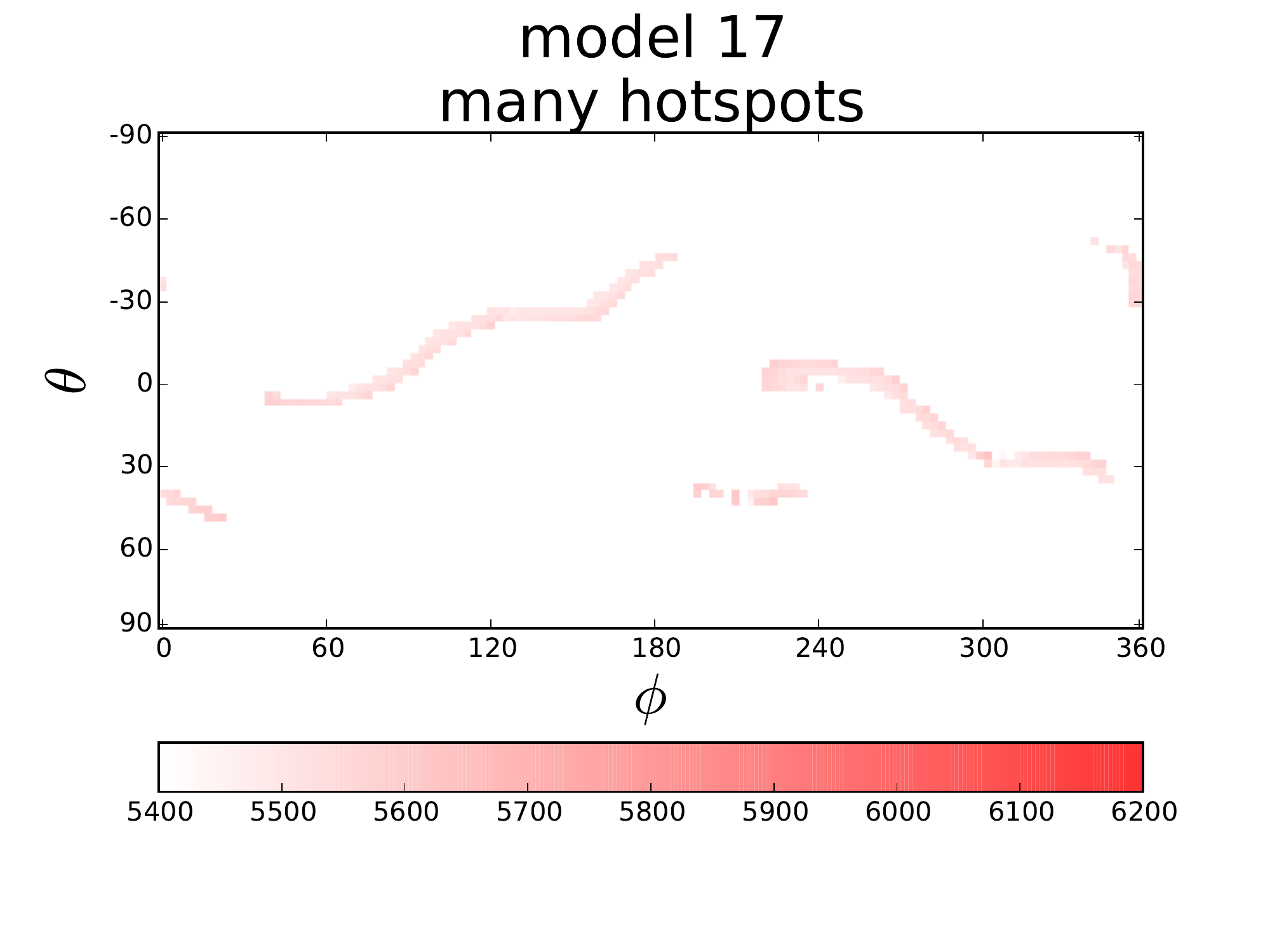}
\caption{\small
Temperature maps of hot spot distributions used for Models 15-17. The x-axis and y-axis represent the stellar longitude ($\phi$) and latitude ($\theta$) in degrees, and the colours indicate different temperatures ranging from 5400 K to 6200 K. Note that the white areas have a much lower temperature of $T_\star$, as indicated in Table \ref{MagneticData}.
}
\label{f:spot_maps}
\end{center}
\end{figure*}

For Models 15-17 we used hot spot maps produced by the isothermal accretion flow described in \citet{gregory07}. We assumed different magnetic field geometries for the three different models, chosen to represent the range of large-scale magnetic topologies observed for accreting PMS stars \citep{gregory12}.  The magnetic fields of PMS stars are observed to correlate with stellar mass and age, hence for each of the Models 15-17 we have considered a different set of stellar parameters (see Table \ref{MagneticData}). Models 15 and 16 had magnetic fields consisting of a slightly tilted dipole component plus a slightly tilted octupole component of different polar field strengths, as listed in Table \ref{MagneticData}.  The tilted magnetic field components are motivated by modeling of spectroscopic observations, which show the large-scale field (i.e. the dipole component) is tilted by $<\sim 20\arcdeg$ in many but not all cases \citep{gregory11}. In Model 15, the dipole and octupole moments were close to an anti-parallel configuration, where the main positive pole of the dipole is close to the main negative pole of the octupole, with the dipole component the dominant field mode.  In Model 16, the octupole component is the dominant field mode, with the dipole and octupole moments close to parallel, where the main positive poles of the dipole and octupole are close to aligned.  These dipole-plus-octupole magnetic field models were analytically constructed and a detailed description of their structure is given in \citet{gregory11}.  Models 17 is a complex magnetic field derived via field extrapolation from a magnetic map of the accreting PMS star V2247 Oph \citep{donati10}.  It is not a true model of that star, but has been adopted here to allow us to consider a magnetic field with an observed degree of complexity.  The dipole component of its multipolar magnetosphere is about 200 G, see Table \ref{MagneticData}.


\begin{deluxetable*}{c c c c c c l}
\tablecolumns{6}
\centering
\tablewidth{0pt}
\tablecaption{Stellar and magnetic field parameters for Models 15-17} 

\tablehead{\colhead{ } & \colhead{$M_\ast (M_{\sun})$} & \colhead{$R_\ast (R_{\sun})$} & \colhead{$T_\ast (K)$} & \colhead{$B_{dipole} (G)$} & \colhead{$B_{octupole} (G)$} }

\startdata
                 
Model 15 & 0.7 & 1.48 & 3989 & 1600 & 400\\ 
Model 16 & 1.3 & 1.94 & 4467 & 800 & 2400\\
Model 17\tablenotemark{1} & 2 & 2.1 & 4955 & 200 & - \\
\enddata

\tablenotetext{1}{Model 17 consists of both the dipole component listed here and a complex component mentioned in the text}
\label{MagneticData}
\end{deluxetable*}

The accretion flow code produced a map of number density of accreted material onto a grid at the stellar surface. By assuming that the material was free-falling along the magnetic lines, we converted this into the accretion rate onto each cell using $\dot{M} = \rho_{*}v_{*}A_{*}$ ($A_{*}$ was the area of the grid cell, $\rho_{*}$ was the density of the material and $v_{*}$ was the infall velocity). This allowed us to calculate the luminosity and temperature of each grid cell on the stellar surface (see Figure \ref{f:spot_maps}). 

 In order to mimic the accretion geometry better for each of these models we also added two disk warps, placed at the phases that the dipole component of the magnetic field was tilted towards, since this part of the field always dominates on larger scales \citep{adams12, johnstone2014}.

In all three of the models, the peaks at each wavelength are misaligned with one another due to the azimuthal misalignment between the hot spots and the warps. Since the main difference between Model 15 and Model 1 is the shape and size of the spots, they have some similarities in terms of the light curve shapes.  The azimuthally extended hot spots of Model 15, however smooth out some of the features, such as the small spike observed in Model 1 at $i=60^\circ$. 
In Model 16 the variability in the infrared and optical appear completely uncorrelated and have a much smaller $\Delta {\rm mag}$ than Model 15. Finally, Model 17 has barely any change in the magnitude, leading us to conclude that the more complex the magnetic field and hotspot structure, the less variability we observe. By looking at Figure \ref{f:spot_maps}, it is clear from the spot map for model 17 that hotspots cover a wide range of latitudes and longitudes across the surface of the star and so the variability is reduced.

\section{Comparison to Observed Lightcurves}

A specific category of young star variables that is garnering increased
attention in ground-based and space-spaced (e.g., CoRoT, Spitzer, K2) datasets
is the ``dipper" category. 
Specifically, the narrow and broad dip stars studied by \cite{stauffer2015} and \cite{mcginnis2015} have optical amplitudes typically $<$0.6 mag and fractional widths up to 1/2 of the period, with the broad-dip stars deeper than the narrow-dip stars.  The models presented here have a similar typical range of $<$0.7 mag and fractional widths up to 0.6 of the period in the V-band.  The great diversity in shapes among both the observed and model lightcurves means that only qualitative rather than detailed quantitative comparisons are meaningful.   We thus suggest that the basic framework presented in this paper of rotationally modulated accretion hotspot illumination of the circumstellar disk can explain some basic lightcurve characteristics, as well as some of their diversity.

A feature of our models when viewed at low inclinations is that the optical and infrared variability is $180^\circ$ out of phase. This feature arises at low inclinations because the surface area of the emitting warm dust is largest when the disk warp is on the far side of the star. For higher inclinations the optical and infrared lightcurves exhibit in-phase variability as described in Section~3.1 and Table~2. For a low inclination system with one or two accretion hotspots the inner disk wall will be brightest 
when illuminated by the hotspot, which occurs when the hotspot is on the far side of the star. However, this phase shift between the optical 
and infrared light curves is rarely seen in real YSOs. Only one of the 162 CTTs in the NGC~2264 YSOVAR monitoring program shows this effect, while it is much more common for the infrared variability either to be in phase with the optical or for the two light curves to show essentially no correlation \citep{cody14}. 
If the YSOVAR
data is interpreted in the context of our hot spot models, the fact that the optical-infrared anti-correlation
is not observed suggests that hot spots are not occulted by the star or the disk warp on the near side,
and so can be observed throughout the stellar rotation period.  This could be attributed  to one or more of:
only moderate viewing inclinations, a combination of inclination and spot latitude effects, or a more complex spot distribution than
we have adopted.

\section{Discussion}

Using our models we can predict percentages of stars in each of the variability categories.  
We assume either a stable (ordered dipole behavior) or unstable disk (chaotic magnetic field, clumpy disk) for all of the stars, and treat these two types of disks as separate cases. The numbers we use in the following paragraphs are estimates from extensive modeling of a grid of inclination angles (e.g., variability is present at 65\arcdeg but dipping at 70\arcdeg).
We will first assume that all of the accretion disks are stable, and that above $i=77^\circ$, 
the wavelengths we are observing would be extincted and the stars will be too faint to detect. 
We will therefore normalize the models over the range $0^\circ \le i \le 77^\circ$. 
Models with $0^\circ \le i \le 20^\circ$ will show little variability, since a high latitude hotspot will be visible throughout 
the entire rotation period. Thus we estimate that about 8\% of stars 
will show no variability. For inclinations $20^\circ \le i \le 67^\circ$ we expect to see some sort of
 periodic variation, so 71\% of stars with stable magnetospheric accretion should show this form of variability. 
 Dippers are likely to show up for $67^\circ \le i \le 77^\circ$, which is 21\% of the stars.
  
 Now we consider the statistics if all of the accretion disks are unstable and therefore have 3-D 
 variations in their disk structure rather than one or two warps.  For $0^\circ \le i \le 50^\circ$ we predict that 
 there will be no variation since with an unstable accretion disk there are not strong hotspots 
 or a pattern of variation from the accretion disk except at high angles of inclination. 
 This means that about 47\% would be non-variable.  For $50^\circ \le i \le 77^\circ$ there will be aperiodic variations, which is about 53\% of the stars.  
   
 Morales Calderon et al. (2011) report that about 70\% of the stars observed were variable. 
 Using this percentage we can try to match our predictions with 
 the observational data. In order to get around 30\% of sources that are non-variable we can estimate that about 50\% 
 of the disks must be stable and 50\% are unstable, giving us 28\% that are not variable. 
 Next we can apply this same 50\% to the rest of the categories to come up with some predictions. 
 Table~4 summarizes this statistical analysis of our models. 
 It is important to note that these results are only for Class II objects and do not include Class I objects that are heavily embedded, or
 spotted weak-lined T-Tauri stars, which are usually categorized as periodic or non-variables.  These statistics include all the main sources of variability, since for Class II objects variability seems dominated by disk-related effects rather than the underlying cool spot rotational modulation, which is undoubtedly there, but not included in our models. In an optical study, \citet{cody14} find that only 3\% of a disk-selected sample showed purely periodic behavior due to spots.

\begin{table}[ht]
\centering  
\caption{Occurrence of YSOVAR classes in our models} 
\begin{tabular}{c c c c} 
\hline                       
\hline                       
 & Stable Disks & Unstable Disks & Total \\ [0.5ex] 
\hline                  
Non-Variable & 8\% & 47\% & 28\% \\ 
Periodic & 71\% & - & 36\%  \\
Periodic Dippers & 21\% & -  & 10\%  \\
Wild & - & 53\% & 26\% \\ [1ex]
\hline 
\end{tabular}
\label{table:predic} 
\end{table}

\vspace{20pt}
\section{Summary}

We have constructed accretion disk models to explain the broad categories of multi-wavelength photometric 
variability observed in the Orion Nebula Cluster and NGC 2264. 
The four main parameters in our models that lead to the different variability are star-spot temperature 
contrast; radius of the inner disk (this determines whether scattering or thermal emission is dominant); 
size and shape of the inner disk warping (the warp presents a different area of the inner disk wall as it 
rotates in and out of view); and system inclination. 
At certain inclinations the 
variability is dominated by occultation of the star by the warped disk. At low inclinations the infrared 
variations are small (the projected area of the inner disk wall is independent of phase) and variability 
is from the spotted star. At high inclinations the mid-IR variability decreases (because thermal emission 
from the inner disk is occulted) and near-IR variability increases due to scattering of light from the hotspots. 

The unprecedented quality of recently available multi-wavelength and
high cadence time series data on young stars now enables detailed
comparisons to magnetospheric accretion models.   Future observations
of e.g. polarization over a rotation period, will further test the models


\acknowledgements

We thank the referee for a careful and thorough report that clarified many points in our manuscript. 


\clearpage

\end{document}